\documentclass[fleqn,usenatbib,useAMS]{mnras}

\usepackage[T1]{fontenc}
\usepackage{ae,aecompl}
\usepackage{graphicx} 
\usepackage{amsmath,amssymb}

\usepackage[british]{babel}
\usepackage{txfonts} 
\newcommand{\degrees}{^{\circ}}         
\title[The Distances of the Galactic Novae]
{The Distances of the Galactic Novae}

\author[\"Ozd\"onmez, A. et al.]{
Aykut \"Ozd\"onmez$^1$\thanks{E-mail: aykut.ozdonmez@ogr.iu.edu.tr}
Tolga G\"uver$^2$,
Antonio Cabrera-Lavers$^{3, 4}$ 
and Tansel Ak$^2$
\\
$^1$Istanbul University, Graduate School of Science and Engineering, Department of Astronomy and Space Sciences, 34116, Beyaz\i t, Istanbul, Turkey\\
$^2$Istanbul  University,  Faculty of Science,  Department  of
Astronomy and Space Sciences, Beyaz\i t, 34119, Istanbul, Turkey \\
$^3$ Instituto de Astrof\'isica de Canarias, V\'ia L\'actea s/n, 38205, La Laguna, Tenerife, Spain\\
$^4$ Departamento de Astrof\'isica, Universidad de La Laguna, 38206, La Laguna, Tenerife, Spain}

\date{Accepted 2016 June 3. Received 2016 June 3; in original form 2016 March 11}
\volume{461}
\pubyear{2016}
\pagerange{1177--1201}

\begin{document}
\label{firstpage}

\maketitle

\begin{abstract}
Utilizing the unique location of red clump giants on 
colour-magnitude diagrams obtained from various near-infrared surveys, we
derived specific reddening-distance relations towards 119 Galactic 
novae for which independent reddening measurements are available. 
Using the derived distance-extinction relation and the independent 
measurements of reddening we calculated the most likely distances for each 
system. We present the details of our distance measurement technique and 
the results of this analysis, which yielded the distances of 73 Galactic 
novae and allowed us to set lower limits on the distances of 46 
systems. We also present the reddening-distance relations 
derived for each nova, which may be useful to analyse the different 
Galactic components present in the line of sight.
\end{abstract}



\section{INTRODUCTION}
Novae are a sub-type of cataclysmic  variables (CVs), in which a white
dwarf  and  a donor  star  form  a  short-period close  binary  system
\citep[see][]{Warner95,BodeE08,  WoudtR14}.    They  are   defined  as
systems with the largest outburst amplitude among other sub-types of
CVs  during which  the  thermonuclear runaway  within a  hydrogen-rich
layer  on the  surface  of  the white  dwarf  leads  to the  explosive
ejection of the nova shell.   Classical nova outbursts can be observed
only once  in the observational history  of a nova, while  with higher
mass transfer rates and more massive white dwarf companions than a typical
nova, recurrent novae show  repeating smaller amplitude nova outbursts
within their lifetimes \citep[see for a review][]{Schaefer10}.

Since novae are exclusively discovered  during   an  outburst  phase,
classification of  these systems  is generally  based on  the observed
properties of this event eg., speed class \citep[e.g.][]{Gaposchkin57}
or  spectral class  \citep{Williams92}.  Alternative  classifications,
based solely on the evolutionary state  of the companion star or
the shape and  the time to decline by 3 mag from the
peak  of  the outburst,  have  also  been proposed  \citep{Strope10,Darnleye12}.  
In addition,  novae have also been  classified as `disc'
or `bulge' novae, on the basis  of their parent stellar population and
location within  the host galaxy  \citep{Dellavalle98}.  Specifically,
the He/N class  novae tend to concentrate close to  the Galactic plane
with a typical  scale height of $\leq100$ pc, whereas  the Fe~\textsc{ii} novae
are  distributed more  homogeneously up  to $z  \leq1000$~pc, or  even
beyond  \citep{Dellavalle98},   where  $z$  represents   the  vertical
distance from the Galactic disc.  For our Galaxy, spatial distribution
analyses of novae predict that they  should be located mostly close to
the  Galactic  plane  with  a typical  scale  height  of  $\sim125$~pc
\citep{Duerbeck81, DellaVella92,  Ozdonmez15}. However,  these studies
lack the necessary statistical significance due to considerable errors
in the distance  measurements.  Measuring the  distances of novae
  in a reliable way can be very  useful in the calculation of both the
  astrophysical parameters and the spatial distribution of these close
  binary systems.

The expansion parallax method allows the  distance to be determined if the
angular  expansion of  the nova  shell can  be resolved  several years
after the  outburst.  However,  application of the  expansion parallax
method runs  into some difficulties.  First,  the expansion velocities
measured  from   the  profiles   of  various  spectral   lines  differ
significantly.   Second,   a  uniform  shell  expansion   has  to  be
assumed. Third, an  uncertainty in estimating the angular  size of the
shell often arises  because of its complex structure.   Despite all of
the uncertainties,  the distance  estimates obtained by  the expansion
parallax method  are believed to  be the most reliable  ones following
after the trigonometric parallax method.  However, the number of novae
with detected  shells is only  about 30 in the entire  history of
their  observations.  Furthermore,  all of  these are  nearby systems,
usually not  farther than  2~kpc \citep{Slavine95,DD00}.  If the
  outburst light  curve of  a nova  is observed,  its distance  can be
  photometrically  calculated  from  the  Maximum  Magnitude  Rate  of
  Decline (MMRD) relations  for Galactic \citep[e.g.][]{Cohen85, DD00}
  and extragalactic nova populations
  \citep[e.g.][]{Capaccioli90,dVL95,Darnleye06}.   However,  a  recent
  study by \citet{Kasliwale11}  has discovered a sample  of faint and
  fast extragalactic novae that caused the authors to question whether
the  MMRD relation is  justified at all, and they  also demonstrated that
  the MMRD relation does not work for recurrent novae.

In this study,  we aim to measure  the distances of a  large number of
Galactic novae,  for which independent extinction  measurements exist,
in a  systematic manner using the  red clump giants (RCs)  in the same
line of  sight as a standard candle.  The  distance of each nova
was  determined  by  comparing  the  specific  extinction--distance
relation for the line  of sight of the nova derived  from the RCs, and
its independent extinction measurement.

Section \ref{extsec} presents an overview  of the various methods used
to determine the amount of  extinction for Galactic novae, emphasizing
their advantages and disadvantages in  the determination of the actual
amount of interstellar matter.  In Section \ref{methodsec} we describe
our method in detail, while we  present the results of the application
of our method  to a large number of novae  in Section \ref{disres}.  A
review of  our method and  a discussion of the  possible uncertainties
associated with  it are given  in Section \ref{concsec} where  we also
discuss  some of  the  implications  of our  results.  Finally, as  an
appendix, we present a short  review of the reddening measurements and
earlier distance determinations for each individual object.
\section{Reddening estimates of Galactic novae}
\label{extsec}

Our aim  is to determine the  distances of a sample  of Galactic novae
following a systematic  approach. In this context,  an independent and
reliable measurement  of interstellar  reddening towards each  nova in
question is needed.   In this section, we discuss some  of the methods
used to  infer the interstellar  reddening towards a certain  nova and
present the value used in this study together with a comparison of the
measurements for each nova in Appendix \ref{appAsec}.

There are a number of methods,  all with varying accuracy, to estimate
the  interstellar  extinction towards  a  Galactic  nova.  For  stellar
studies there  is a rich set  of spectroscopic methods, which  lead to
the estimation  of  the interstellar  extinction.  For  example
$E(B-V)$ is best determined by comparing line fluxes of the source and
an   unreddened  standard   object,   which   have  nearly   identical
spectroscopic features  to the  source in question.   However, neither
this method nor  the continuum fitting method based on 
stellar atmospheric  models can be  applied to  a nova because  of the
fact that the models are still  quite rudimentary and the lack of such
reference  objects.  Thus,  indirect  methods  are usually  preferred.
These  methods often  require either  high--resolution  spectroscopy to
reveal  interstellar  lines  \citep[e.g., Na ~\textsc{i}  ($\lambda5890.0$  and
  $\lambda5895.9$) and  K~\textsc{i}  ($\lambda7699.0$), the line  strengths of
  optical and near-IR spectroscopy of O~\textsc{i} lines][]{MunariZ97, Rudye89}
or  other lines,  which are  known to  be correlated  with the  colour
excess.  There are a few difficulties  in using spectral lines for the
determination of the interstellar  reddening. First, the spectral line
widths of  novae are large  and the  line structure is  complex.  Line
profiles are occasionally asymmetric,  making it difficult to separate
an individual spectral line from  blended features.  Second, it is not
always possible to examine the same spectral line for all novae, since
the spectral  evolution of  each nova  is unpredictable  and peculiar.
Another reddening estimation method is based on the 2200 {\AA} feature,
which  is  characterized  by  the presence  of  a  broad  interstellar
absorption bump in the spectral  energy distribution (SED).  Since the
ultraviolet (UV) band is  very sensitive to the effects of  the interstellar medium,
even a  small amount of reddening  can be derived by  modelling the SED
with an $E(B-V)$ that accounts  for the extinction bump.  However, low
sensitivity of the  spectrographs in the UV wavelength  range and high
noise complicate the estimation of the interstellar extinction.

On the other  hand, the colour index  of a nova at  a certain outburst
phase  can  be used  to  also  estimate the  interstellar  extinction,
assuming  that all  novae  have  similar colour  indices  at the  same
outburst  phase.   \citet{vandenbergY87}  derived  general  trends  of
colour evolution in  the outburst light curves of  novae, and inferred
the intrinsic  colour of  novae as $(B-V)_0=0.23\pm0.06$~mag  when the
brightness     of    a     nova     reaches     its    maximum     and
$(B-V)_0=-0.02\pm0.04$~mag when the nova faded 2 mag from the
outburst  maximum,  where the  subscript `0'  denotes the  unreddened
colour.   Similarly,   \citet{Miroshnichenko88}  suggested   that  the
intrinsic  colour is  $(B-V)_0=-0.11\pm0.02$~mag at  the stabilization
stage, which occurs  when the colour evolution of  the nova stabilizes
soon  after the  optical  maximum.   Recently, \citet{Hachisu14}  have
improved the colour-colour evolution  of novae using revised distances
and extinctions,  and compared the reddening  estimates obtained using
the intrinsic colours given  above.  \citet{Hachisu14} showed that the
intrinsic colour methods  often show large deviations, and  the one at
the stabilization stage seems to be more useful.  However, since novae
in general are peculiar non-stationary objects, and a nova outburst is
an  unpredictable  event,  it  is always  possible  to  miss  an
observation that must be done at a certain phase.

\begin{figure}
	\centering
	\includegraphics[width=0.35\textwidth]{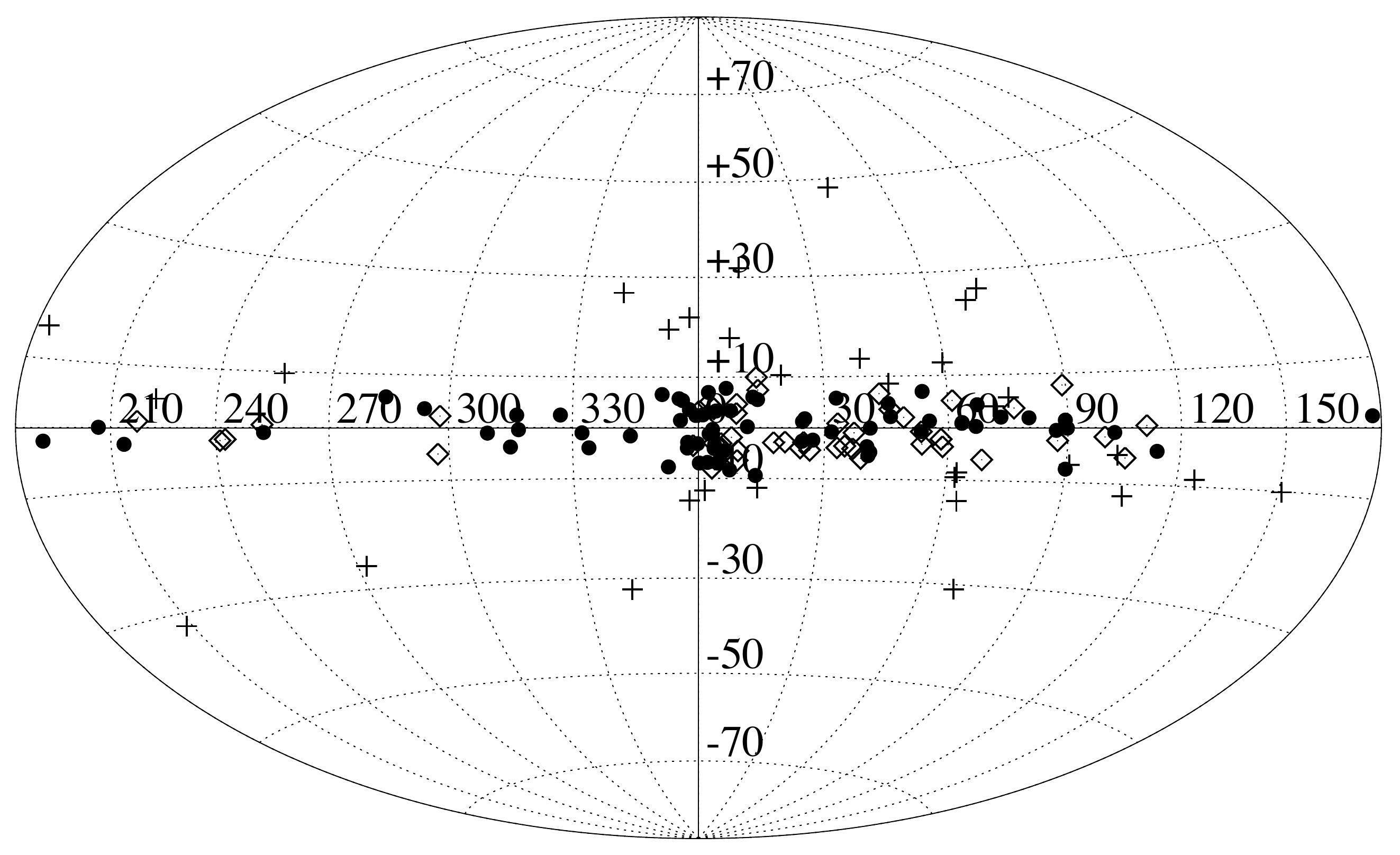}
	\caption{Galactic  coordinates of  novae which  have reddening
          estimates. ($\bullet$), ($\vartriangle$)  and ($+$) indicate
          the ones  for which  the distances  could be  obtained, only
          lower limits could be calculated, and the ones for which the
          distance could not  be calculated, respectively. Estimations
          were explained in Section  \ref{disres}.}
	\label{fig:fig1.1}
	\end {figure}

We  collected  all  the interstellar  reddening  estimates,  published
before 2016 April, for 152  Galactic novae from the  literature. Note
that, the  estimates based on  the use of interstellar  reddening maps
were  ignored.  Galactic  coordinates of  these systems  are shown  in
Fig. \ref{fig:fig1.1}.  For the  distance calculation, we preferred to
use the  reddening values obtained  from spectra, especially  the ones
measured through  the use of uncontaminated  lines, e.g. Na ~\textsc{i},  K~\textsc{i} or
Hydrogen column  density, 2200  \AA\ feature  and finally  line
ratios.  Reddening values obtained from  photometry were used if there
is  no  other  reddening  value available  from  spectroscopy.   Short
discussions of the adopted reddening estimates of each individual nova
are given in Appendix~\ref{appAsec}

\section{Method}%
\label{methodsec}
\subsection{Red Clump Stars As A Standard Candle}%

Just  like a  normal  horizontal branch  star the  RC  stars are  core
helium-burning  giants.   They have  long  been  proposed as  standard
candles \citep{PacS98,Lopere02},  owing to their  compact distribution
in  colour-magnitude  diagrams (CMDs)  of  open  clusters as  well  as
globular  clusters. They  are  well separated  from the  main-sequence
stars and red giants.   Location of the RC stars in  a CMD presents an
opportunity to obtain  a measurement of their  mean absolute magnitude
and intrinsic colour.   Furthermore, and especially in  a near-IR CMD,
dependence of the  location of these stars on the  metallicity is very
weak. Thus, it  is possible to use these stars  as tracers of Galactic
reddening-distance relation in a given line of sight, accounting for a
small  dispersion of  their absolute  magnitude and  intrinsic colour.
Such a relation can then be used to infer the distance of a particular
object, in the same field of  view, for which the extinction is known.
A very similar technique has been successfully applied before to
several X-ray sources \citep{DurantV06,Guvere10}.

To  use the  mean  absolute magnitudes  of the  RC  stars in  distance
estimation,  their  absolute magnitudes  and  intrinsic  colours in  a
particular photometric system  are expected to be  free of metallicity
and age.  Considering the $K_\text{s}$ and $J$ bands of the Two Micron All-Sky Surver (2MASS) photometric
system, which are  used in this study, numerical  studies suggest that
the  infrared  magnitudes, especially  $K_s$ band,  have only  a
small  dependence   on  metallicity   and  age   both  observationally
\citep[e.g.][]{GrocholskiS02,Pietrzynetal03}     and     theoretically
\citep{SalarisG02}.  The intrinsic  colour $(J -K )_0$  is affected by
these  parameters  due to  the  dependence  of  the  $J$ band  on  the
metallicity and age.  For our case, this effect is negligible as there
is   no   large   metallicity    gradient   in   the   Galactic   disc
\citep{IbataG95a,IbataG95b,Sarajedinietal95}.   Here,  we  adopted  an
absolute  magnitude of  $M_{K_\text{s}}=-1.595\pm0.025$~mag  and an  intrinsic
colour of  $(J-K_0)=0.625\pm0.03$~mag from  \citet{Yaze13} for  the RC
population based  on the re-reduced  {\it Hipparcos} RC data using  the same
methodology as  in \cite{Alves00}  and \cite{Laneye12}.   These values
are not only the most recent ones, but also compatible with the former
ones;  cf.   $M_K=-1.61\pm0.03$  mag  \citep{Alves00}  from  {\it Hipparcos}
sample,  $M_{K_\text{s}}  =-1.62  \pm0.06$~mag   from  14  open  clusters  and
$(J-K_\text{s})_0=0.597\pm0.033$~mag   for   NGC2158   \citep{GrocholskiS02},
$M_{K_\text{s}}  =  -1.57\pm0.05$~mag  for 24  open  clusters  \citep{vanG07},
$M_{K_\text{s}}  =   -1.54\pm0.03$~mag  \citep{Groenewegen08}  based   on  the
re-reduced   {\it Hipparcos}   catalogue   \citep{vLeeuwen07},   $M_{K_\text{s}}   =
-1.613\pm0.015$~mag  and $(J-K_\text{s})_0=0.63\pm0.2$~mag  for the RC
stars in the solar neighbourhood \citep{Laneye12}.

\subsection{The Photometric Data}%
 \label{pdsec}
In order to use the specific extinction-distance relation derived from
the RC stars in a Galactic field including a particular Galactic nova,
the $(J-K_s,  K_s)$ CMD  should be constructed  first.  For this
  purpose,  we  used  several   near-IR  surveys,  namely,  (i)
2MASS   \citep{Skrutskiee06},  (ii)
Visible and Infrared Survey Telescope for Astronomy-Vista Variables in
the Via Lactea  survey \citep[VISTA-VVV;][DR2]{Saitoetal12}, and (iii)
the  United Kingdom  Infrared  Deep Sky  Survey-Galactic Plane  Survey
\citep[UKIDSS-GPS;][]{Lucase08}. All of  these surveys complement each
other regarding the completeness limits  and sky coverage, which is an
advantage that  allows us not only  to trace the RC  stars through the
stellar fields,  but also to study  Galactic novae in a  wide-range of
the distances.  For example, 2MASS  extends the study further from the
Galactic plane owing  to its observing policy covering  the whole sky,
whereas     UKIDSS-GPS    is     only    mapping     $-0.5\degrees\leq
l\leq110\degrees$,   $|b|<5\degrees$  in   the  Galactic   plane,  and
VISTA-VVV observes the Galactic bulge in range $-10\degrees \leq l \leq
+10\degrees$, $-10 \degrees \leq b \leq +5 \degrees$ and the Galactic disc
in range  $295 \degrees  \leq l  \leq 350  \degrees$, $-2\degrees  \leq b  \leq
+2\degrees$.   Here,  $l$  and  $b$ are  the  Galactic  longitude  and
latitude, respectively. The lower limit of magnitudes in the VISTA and
UKIDSS $K$  bands are about 18.0~mag  for most of the  stellar fields,
while it reaches to $\sim16.5$~mag for the crowded fields, which
are approximately  4 mag fainter  than 2MASS.  On  the other
hand, 2MASS photometry allows us to  determine the nearest RC stars in
the CMDs  up to $\sim8$ mag,  which is 3 mag  lower than the
saturation limit of $\sim11$ mag in VISTA and UKIDSS $K$ bands.

Since the  photometric bands  of the UKIDSS  survey are  not based  on the
2MASS zero-points, we converted them to the 2MASS magnitudes of the RC
stars using the transformation equations given by \citet{Wegge15}:

	\begin{equation}\label{Eq:1.1}
	\begin{aligned} 
		J_2 &= J_\text{U} -0.02(J_\text{U} -H_\text{U} )-0.03 \\
		H_2 &=H_\text{U} +0.12(H_\text{U} -K_\text{U}) \\
		K_2 &= K_\text{U} +0.02(H_\text{U}-K_\text{U} )+0.01,
	\end{aligned}
	\end{equation}\\*

where the  subscript $2$  refers to  the 2MASS system  and subscript U refers  to the
UKIDSS.

Using $J$  and $K$ magnitudes  from each survey, when  available, CMDs
were  created to  estimate  the  distance of  a  given  nova from  the
reddening-distance relation.  In our analyses, we preferred to use VVV
data, switching  to UKIDSS and  finally to 2MASS.   However, for
some  of the  directions, it  was necessary  to combine  two or  three
catalogues  by  cross-matching  the   object  coordinates.   For  this
purpose,  we  used a  matching  radius  of  1arcsec and  limit  the
magnitude and  colour differences to  $|K| \leq 0.5$ mag  and $|J-K_\text{s}|
\leq 0.2$ mag, respectively.
\subsection{Selection of the RC stars and the reddening-distance relation}%
 \label{rdsec}
 
Positions of  the RC  stars on  a CMD  depend on  their absolute
  magnitude, intrinsic colour, distance and extinction. Since absolute
  magnitudes and intrinsic  colours of RC stars are  similar, a change
  only in  the extinction  causes a  horizontal displacement  of their
  position while a  vertical displacement of the position  is a result
  of distance or extinction change.   Therefore, the extinction can be
  determined by  comparing intrinsic colour and  mean observed colour.
  Using this  information and  the mean absolute  magnitude of  the RC
  stars, the distance  can be derived from the  Pogson's equation.  On
  the contrary,  in a region  of the  Galaxy with low  extinction they
  trace  almost a  vertical line  on the  CMD throughout  the line  of
  sight.
  
In order  to determine the approximate region in which  the RC
stars  are   located  in  an   observational  CMD,  we   used  \textsc{galaxia}
\citep{Sharma11}. This is  a code to generate synthetic  models of the
Galaxy. The code constructs a catalogue  of stars based on the specified
colour and  magnitude limits and  a coordinate set.   The interstellar
extinction  (or reddening)  at  any given  distance  and the  Galactic
coordinates of a star  can be calculated by a model  of the density of
interstellar material \citep[for details see ][]{Sharma11}]:

\begin{equation}\label{Eq:1.4}
E_d(B-V)=E_{\infty} (B-V)  \frac{\int_0^r \! \rho(s)\text{d}s}{\int_0^{\infty} \! \rho(s)\text{d}s},
\end{equation} \\

\begin{figure}
	\centering
	\includegraphics[width=0.44\textwidth]{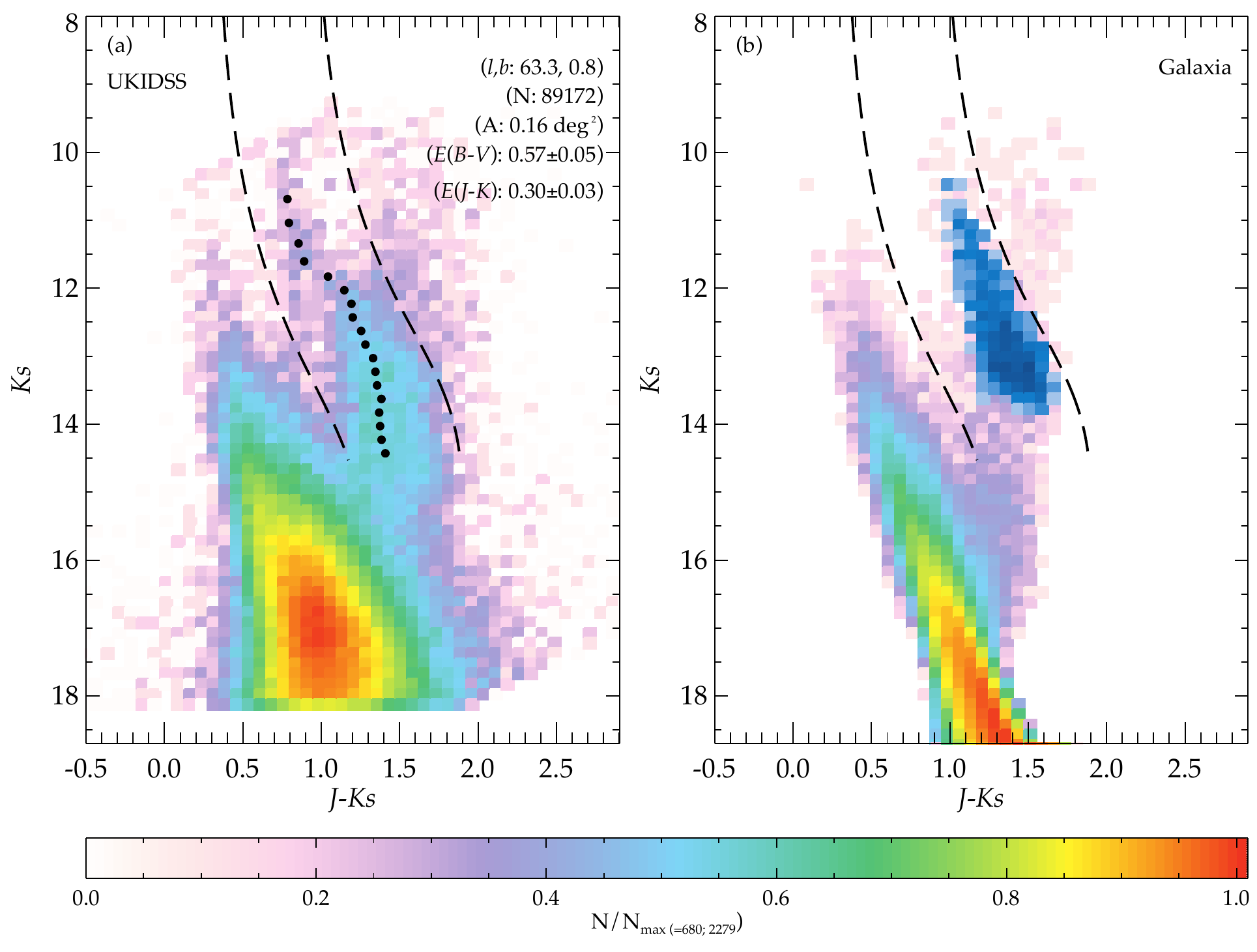}
	\caption{(a)  Observational (b)  \textsc{galaxia}  CMDs of  LV Vul  for
          $0.4\times0.4$ deg$^2$ line of sight. Black dashed lines are
          boundary lines  for selections of  the RC stars;  black dots
          show $J-K_{s}$ colours of RC  stars; \textit{N} is the total number of
          stars  in the  area,  \textit{A} is  the solid  angle,  $l,\: b$  are
          Galactic coordinates. The reddening  estimate of the nova is
          also given.  Colour scale  is shown  below the  figure. Note
          that: maximum number  of stars for each bin is  shown by red
          colour.}
	\label{Fig:1.2}
	\end {figure}

Following  the results  of \citet{Arce1999}  and \citet{Schlafly2011},
colour  excess at  infinite  distance given  by \citet{Schlegel98}  is
reduced by  a  factor depending  on  interstellar
reddening  and  the line  of  sight.   We  obtained  the RC  trace  by
evaluating  the  interstellar  extinction  in steps  of  5~pc  of  the
distance  using $A_J=0.871  \times  E_d(B-V),  \: A_{K_\text{s}}=0.346  \times
E_d(B-V)$ \citep{RiekeL85}, and placed  them on both observational and
theoretical CMDs.  Boundary lines, which limit the region where the RC
stars are  located on CMDs, were  derived by shifting the  RC trace in
equal  amounts  for  both   directions  on   the  colour   axis  (Figs.
\ref{Fig:1.2}a and b).   The width  of the  band limited  by the  boundary
lines was determined  after a number of trials to  maximize the number
of RC stars while avoiding stars of other spectral types.

Once  a broad  region  in which  the  RC stars  should  be located  is
identified, magnitude  intervals for  the $K_s$  band, from  which the
reddening will be  calculated, can be determined.  The  lower limit of
$K_s$ magnitude interval was derived following the RC build-up.  Then,
the  magnitude intervals  were derived  in steps  of a  given distance
(usually 300 or 400~pc) from the  lower limit.  However, if the number
of stars in a particular magnitude interval was less than 20, then the
range  was extended  to larger  magnitudes until  the number  of stars
became 20, and then new magnitude  intervals were derived in steps of
corresponding  distance from  the  re-constructed  limit.  Since,  the
margin between  magnitude intervals decreases at  larger distances, we
set a lower  limit of about 0.2-0.3~mag on the  range of the magnitude
intervals.

Following the summarized method, histograms  for the colour indices of
stars  in each  magnitude interval  can be  obtained. A  peak in  this
histogram would  mostly reveal the presence  of the RC and  some giant
stars,  that  dominate  the   distribution  of  the  background  stars
(generally  represented by  a second--order  polynomial), because  the
colour difference  between the  RC and  RG stars  is very
small.  Note that the same holds also for the temperatures and surface
gravities of these two  sub-samples.  Thus, we expect  a possible
contamination by the RG stars in  our study. The effects of this
  possible  contamination   will  be  also  considered   in  following
  paragraph.  Fitting  the resulting histograms with  a Gaussian, the
$J-K_\text{s}$  colours for  the  RC stars  can be  derived,  along with  its
apparent width, which is defined through the standard deviation of the
Gaussian  fit.   This  model  is  used and  defined  by  many  authors
\citep[e.g.][and  references  therein]{Staneke95,Cabrerae08,Natafe10},
as follows:
\begin{equation}\label{Eq:1.6}
	\begin{aligned}
N(m) =& \frac{N_\text{RC}}{\sigma_\text{RC}\times \sqrt{2}\pi}  \exp\left[-\frac{1}{2}\left(\frac{\mu-\mu_\text{RC})}{\sigma_\text{RC}}\right)^2\right]\\
& +A\ \exp\left[B\ (\mu-\mu_\text{RC})\right]
	\end{aligned}
\end{equation}\\*
Here, $N_\text{RC}$  is  the  number of  the  RC  stars within  each
magnitude interval, $\mu_\text{RC}$ is the  centre of the fitted curve that
is in  fact the estimated  colour, $\sigma_\text{RC}$ is the  dispersion of
the colour.   $A$ and  $B$ fit  an exponential  profile to  the colour
distribution of the  RG stars.  Note that, we prefer  median values of
the  colour  indices  rather  than  using the  Gaussian  fit  for  the
magnitude  intervals which  contain  less than  50  stars.  For  these
intervals, we  calculate the uncertainty of  $(\sigma_{JKm})$ from the
standard deviation of the  colours. Once the statistically significant
colours  and   the  associated   uncertainties  are   calculated,  the
$E(J-K_\text{s})$ colour  excesses were  derived from the  difference between
these values  and the assumed intrinsic  colour of the RC  stars.  The
relation for the total absorption,  $A_{K_\text{s}}= 0.657 \times E(J-K_\text{s})$, is
adopted from \citet{RiekeL85}.  This equation leads to the calculation
of the distance $d_\text{c}$.

Although the  reddening curve can  be calculated by this  process, the
number  of  RC stars  may not be dominant  in a  given  magnitude
interval, leading to  a difference in the extinction  from the general
trend.  A scatter can also be caused by the background distribution of
stars  from various  spectral  types.   In order  correct  for such  a
possible scatter, we utilized the  \textsc{galaxia} extinction curve to fit the
theoretical  curve  with  the  observational curve  obtained  in  this
study. When an  observed extinction measurement at  a certain distance
deviates  from  the \textsc{galaxia}  prediction  and  also from  the  observed
general  trend  by  about  $2\sigma$,  we  re-analysed  the  magnitude
interval in question using one of the following methods: for the $K_\text{s}$
band  magnitude  interval  in  use,   the  colour  index  $J-K_\text{s}$  was
re-calculated:
\begin{enumerate}
  \item from the peak of the $J-K$ histogram.
  \item from the median of the $J-K$ histogram.
  \item using two Gaussians to fit the $J-K$ histogram distribution.
  \item using two Gaussians plus  a polynomial function model  to fit
    the $J-K$ histogram distribution.
\end{enumerate}

If the new colour excess value  shows a better agreement, it was used.
Otherwise,  we   just  left   the  original  result   from  equation
(\ref{Eq:1.6}).  Note  that such a  correction was only  performed
for individual reddening  measurements and was not used  to modify the
inferred  trend  as  found  observationally  from  RCs.   Finally,  we
obtained the reddening-distance relation for  the line of sight of the
nova in question.

\begin{figure}
	\centering
	\includegraphics[width=0.44\textwidth]{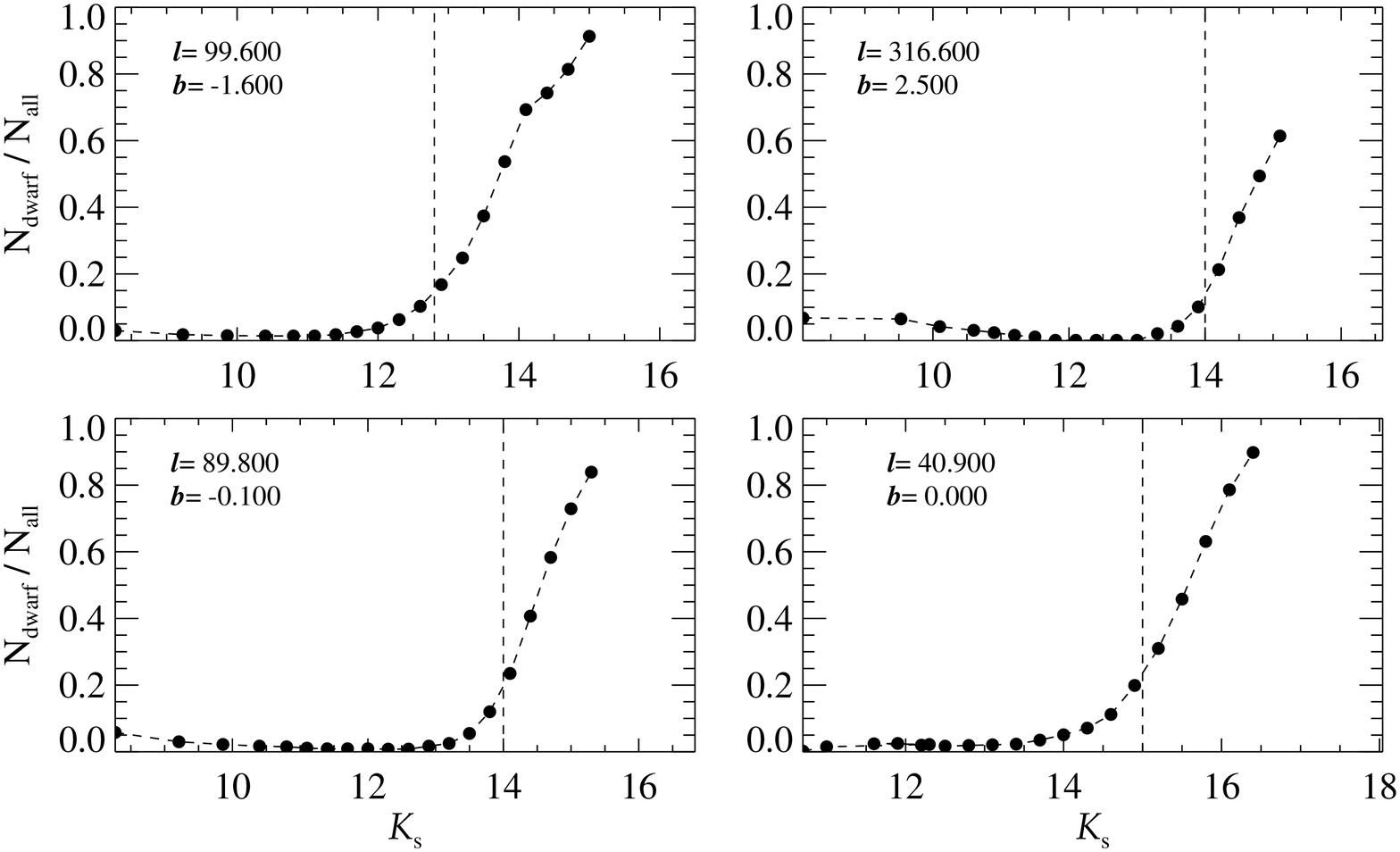}
	\caption{Rate  of dwarf  contamination  towards four  different
          directions.  Dashed  line represents the upper  limit of the
          magnitude interval.}
	\label{Fig:1.3}
	\end {figure}

The  contamination of the  RC sample  caused by dwarf  stars must
  also  be  considered.   The  selection method  used  here  can  not
separate  the  dwarf  and   giant  populations.   In  order  to
understand the influence of  the dwarf/giant contamination, individual
boundary lines were placed on the  CMDs obtained from the \textsc{galaxia} code
(Fig. \ref{Fig:1.2}b) and  predicted ratio between the  dwarfs and the
giants plus  dwarfs were  calculated using  the stellar  parameters in
\citet{Sharma11}: $T_\text{eff}  > 5000 \  \text{K}, \:  \log g  >3.3 $  for dwarfs,
$T_\text{eff} \leq 5000 \  \text{K}, \: \log g \leq 3.3$  for RG+RC stars.  Examples
are given in  Fig. \ref{Fig:1.3} for four different  directions in the
Galaxy.   Note that  the first  points in  Fig. \ref{Fig:1.3}  are not
statistically important  due to lack of  stars.  It is clear  that the
dwarf contribution begins to  be important towards fainter magnitudes,
especially  for magnitudes  fainter than  $K_\text{s}=13$-~14 mag.   The dwarf
contamination for regions  close to the Galactic plane  is expected to
be  smaller  (in  other  words,   its  efficiency  begins  at  fainter
magnitudes), since higher reddening in this region will shift the
  giants to  redder colours. Meanwhile  the dwarfs which  have similar
  magnitudes  as giants  will remain  in the  blue part  of the  CMDs,
  because the effect  of reddening on their colours is  lesser. As a
result,  the dwarf  and giant  stars can  be separated.   Therefore we
considered this  effect in  the calculation of  the reddening-distance
relation for each Galactic nova, and  set an upper limit the magnitude
interval  so  that  the  dwarf  contamination  is  lesser  than
$\sim30\%$.
	
The uncertainty in  the extinction was estimated  from the uncertainty
of    the   mean    colour    derived   from    the   Gaussian    fit,
$(\sigma_{JKm}=\sigma_\text{RC}/(1.52   \times  N_\text{RC})$.    However,  this
uncertainty is only  an indicator of the statistical  error. The total
uncertainty in the  extinction was determined taking  into account the
uncertainty of the intrinsic colour of RC stars, $(J-K)_0=0.03$,
and      adding      both       uncertainties      in      quadrature:
$\sigma_{JK}^2=\sigma_{JKm}^2 + \sigma_{JK0}^2$.

\subsection{Determination of the Distances}
The  distances of  the  Galactic novae  were  estimated comparing  the
reddening-distance  relations  obtained from  the  RC  stars with  the
colour excesses $ E(J - K_\text{s}) = 0.524 \times E(B - V)$ \citep{RiekeL85}
for  the  novae.  As  mentioned  in  Section~\ref{extsec}, the  colour
excesses  of  novae  were  compiled   from  the  literature.   In  the
comparison, we  preferred to use  the probability distribution  of the
extinction,  which  is denoted  as  $P_\text{nova}(EJK)$  for the  nova  in
question  and the  one  obtained  for the  RC  stars for  each
magnitude   bin   as   $P_\text{RC}(EJK|D)$.   These   distributions   were
represented by Gaussian functions.  Since there is no prior either for
the   extinction   or   the  distance,   the   probability   functions
$P_\text{RC}(EJK|D)$ and  $P_\text{RC} (D|EJK)$  were assumed  to be  equal. The
total probability distribution was  calculated taking into account the
product  of  these  two   independent  probability  distributions  and
integrating them over the reddening:
  
	\begin{equation}\label{Eq:1.12}
	P(D) =\int P_\text{nova}(EJK)\:\: P_\text{RC}(D|EJK) \:\: \mathrm{d}EJK.
 	\end{equation}\\

\begin{figure}
	\centering
	\includegraphics[width=0.46\textwidth]{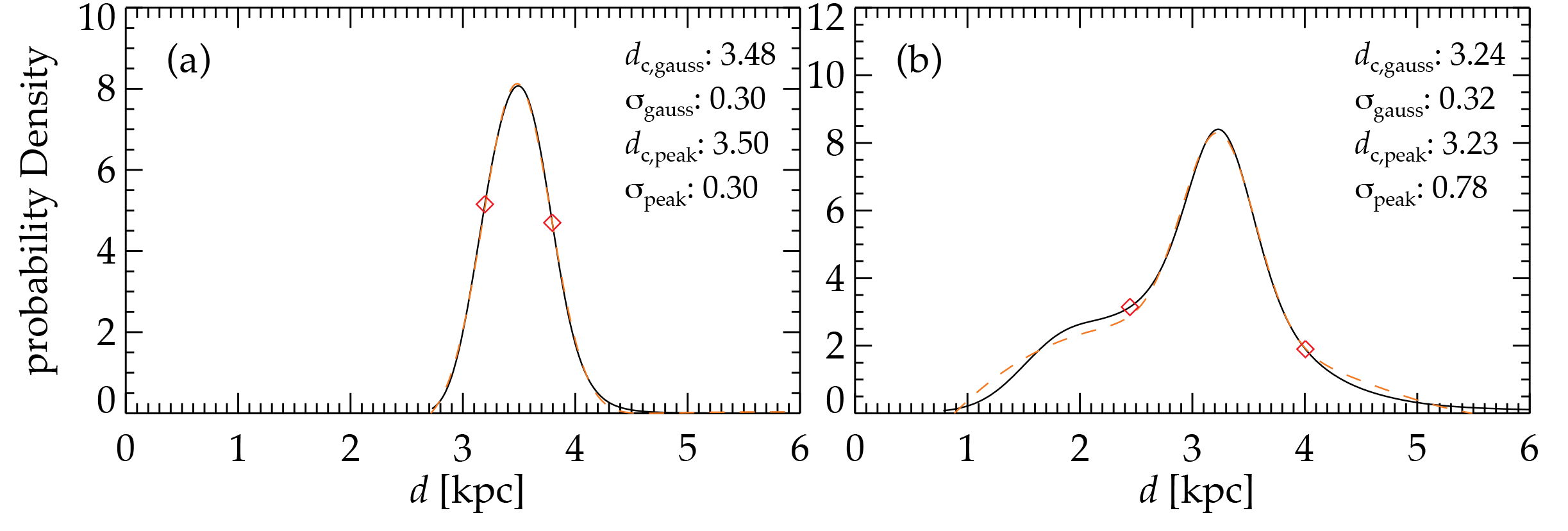}
	\caption{Probability distribution over  distance to the source
          (a) V407  Cyg and (b)  V496 Sct. Red dashed  line represents
          the best-fitting  Gaussian function  and diamond  symbols are
          the distances  with uncertainties calculated from  the integration
          method.}
	\label{Fig:1.4}
	\end {figure}

We   solved  this   integral  and   obtained  the   final  probability
distribution of the distance (Fig.  \ref{Fig:1.4}) of a given nova. In
order  to find  the most  likely distance  from this  distribution, we
smoothed the distribution using a spline function and took the maximum
value.  We fitted the distribution  with a Gaussian curve to determine
the  uncertainty of  our  measurement together  with  the most  likely
distance  value, if  the distribution  had a  Gaussian profile.   
However, the resulting distribution can not always be represented by
  a simple Gaussian  curve.  In these cases, we  calculated the total
area  under  the probability  distribution,  normalized  it to  1  and
determined the range, from the most likely value, in which the area is
equal to the 0.68 times the total. An example for this is shown
in Fig. \ref{Fig:1.4}.

\section{Distances of the Galactic Novae}
 \label{disres}
\begin{figure}
	\centering
	\includegraphics[width=0.46\textwidth]{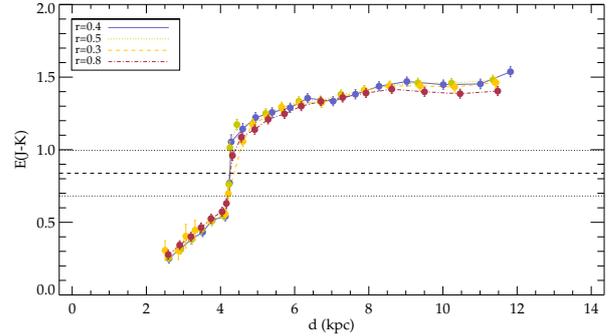}
	\caption{The  reddening-distance  relations   through  WY  Sge
          obtained using four different radii.}
            	\label{Fig:wysge}
	\end{figure}

       In the  application  of the  above summarized  procedure,
         using a  large number of  stars for  each field is  needed in
         order  to  both reduce  the  statistical  uncertainty of  the
         colour  measurement and  to minimize  the contamination  from
         stars of  different spectral  types. However,  increasing the
         number of stars also means increasing the area of the line of
         sight to be  considered, which reduces the  resolution of the
         reddening since the method only  provides the mean in a given
         field of view.  Therefore, we need to maximize the resolution
         while keeping the number of stars  at a minimum.  In order to
         ensure  that  there  are  no significant  variations  in  the
         reddening  curve  through  the  line of  sight,  we  obtained
         reddening  - distance  relations  using  fields with  varying
         sizes towards each nova.  An example case is shown for WY Sge
         in Fig.  \ref{Fig:wysge} where reddening - distance relations are
         obtained using four different radii  of 0.3, 0.4, 0.5 and 0.8
         deg.  The resulting  relations imply that there  is no strong
         variation of reddening in those areas centred on WY Sge.
 
 Using the method detailed in Section \ref{methodsec}, we attempted to
 measure the distances  of 152 Galactic novae in  our initial catalogue,
 which have independent reddening measurements. The reddening-distance
 relations produced from  the RC stars are also useful  to analyse the
 different Galactic components present in  the line of sight since the
 shape  of the  extinction curve  will  be directly  related to  these
 components.
 
The bulk  of absorbing interstellar  matter is concentrated  along the
Galactic plane at  low Galactic latitudes, while  for higher latitudes
the slope of  the reddening curve is shallower since  the dust density
decreases rapidly. QV~Vul or  LW~Ser (Fig. \ref{Fig:A1}) at latitudes
higher than $6\degrees$ is clear examples  of the latter as the majority of
the  dust is  concentrated in  the first  1-2~kpc in  the thin  disc
component.  Further away from the plane, there is no additional
extinction, and  the reddening-distance relation becomes more or
less flat. On the other hand,  for pure disc fields, for example in
the case  of V1500~Cyg  (Fig. \ref{Fig:A1}), the  extinction increases
steadily  along the line of  sight up to the  detection limit or
the end of the disc --whichever comes first-- is reached.

Directions where  other Galactic  components apart  from the  disc are
present, are clearly identifiable from  the results presented here. In
this  regard, the  amount  of  dust concentrated  in  the spiral  arms
translates  in sharp  increases  in  reddening as  a function  of
  distance once the location of an arm is crossed.  This can be
seen  in  the   cases  of  LV~Vul  (Fig.    \ref{Fig:A1})  and  WY~Sge
(Fig. \ref{Fig:A3}) thanks to their relationships with the Sagittarius
arm. The  dust associated with  the arm produces an  extra extinction,
which increases by 0.5$-$0.6~mag in a  few parsecs as the arm is
crossed (at  a distance of around  4~kpc along the line  of sight) and
produces  a recognizable  feature in  the plots.   When moving  to the
inner Galaxy, the  bulge component is the most  prominent feature, and
it  dominates the  reddening-distance  plots producing  a steady 
increase  in extinction  up  to  the distance  where  the bulge  is
reached (as it is shown for example  in the case of V4643 Sgr and V732
Sgr in Fig. \ref{Fig:A3}).

\begin{figure}
	\centering
	\includegraphics[width=0.44\textwidth]{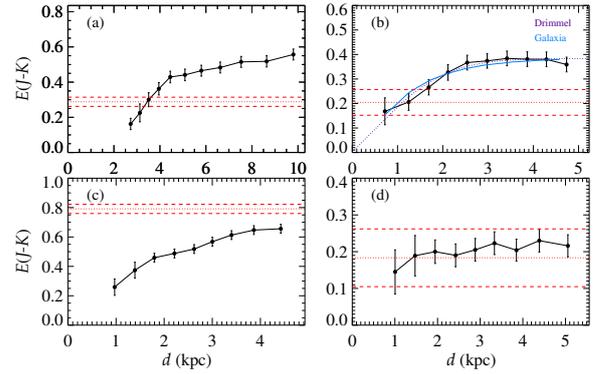}
	\caption{Reddening-distance relations for  four scenarios: (a)
          the distance can be clearly determined; (b) the distance can
          be  determined with  getting help  from models;  (c) only  a
          lower limit  for the distance  can be obtained; and  (d) the
          distance can not be measured.}
	\label{Fig:1.5}
	\end {figure}

 The uncertainties in  the distances of novae found from  the RC stars
 depend on how  the interstellar reddening increases  with distance in
 combination with  the uncertainties  associated with  the independent
 reddening measurements.  For directions where the interstellar medium
 is relatively sparse and therefore the distance-reddening relation is
 not steep  enough, the method  outlined here is not  effective.  Also
 for the novae, for which the reddening measurements already have very
 large  uncertainties,  a  reliable   distance  determination  is  not
 possible  (Fig.  \ref{Fig:1.5}d).   For these  reasons, we  could not
 estimate the distances of 33 objects in our catalogue. These
   novae  are listed  in  Appendix \ref{notcalc}  with a  brief
   information  on  why  their  distances could  not  be  calculated.
 Within   the  remaining   sample,  we   have  seen   three  different
 possibilities.   The first  one is  the case  where we  were able  to
 obtain a reliable distance (Fig. \ref{Fig:1.5}a).  In such cases, the
 distances were  accurately calculated  assuming that  the independent
 reddening  reflects the  correct amount  of the  interstellar matter.
 The  second case  occurred when  the reddening  estimate of  the nova
 corresponds  to   the  first   or  the last  point   in  the
 reddening-distance relation  (Fig. \ref{Fig:1.5}b).  In  these cases,
 we  utilized  three  dimensional  extinction model  derived  by
   \citet{Drimext}, and also the  \textsc{galaxia} code as an independent
   check. We  concluded  that  the  distance  is  reliable  if  our
 reddening evolution  shows the same  trend with Drimmel's  map and/or
 \textsc{galaxia}.  Otherwise,  the distance  calculated from the  first points
 in  the reddening-distance relation will  remain uncertain, and
 only an upper limit of the distance can be given.  The worst scenario
 is the final  case when the reddening estimate of  a nova exceeds the
 lower/upper limit  of the corresponding reddening  curve.  An example
 for such a  case is given in Fig. \ref{Fig:1.5}(c).   In such cases, it
 is  impossible to  predict how  the trend  in the  reddening-distance
 curve proceeds, so we are only able to give a lower or an upper limit
 for the distance of the nova based  on the first or the last point in
 reddening-distance relations.   Taking into account all  these cases,
 we were able to constrain the distances of 119 Galactic novae.  Their
 derived   reddening-distance   relations   are  given   in   Appendix
 \ref{appBsec}.

\begin{figure}
	\centering
	\includegraphics[width=0.34\textwidth]{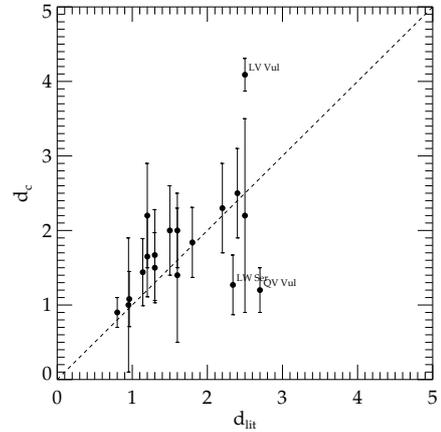}
	\caption{Comparison of  our distances  with those  estimated
          from the expansion parallaxes.}
	\label{Fig:1.6}
	\end {figure}

We compared the distances of  18 Galactic novae estimated in this
  study with those estimated using  the expansion parallax method (see
  Appendix  A) and  listed them  in Table  \ref{table:exp}.  Although
some problems exist in the determination of expansion parallaxes, it
is  the probably  the  most reliable  method  after the  trigonometric
parallax  method.   From  the  comparison  (Fig.   \ref{Fig:1.6}),  we
calculated  the  median  of   the  differences  between  our  distance
estimates and  expansion parallaxes to  be $\Delta d=0.11$~kpc  with a
standard deviation of $\sigma_{\Delta  d}=0.7$~kpc.  These results are
consistent with the median error of the calculated distances, 0.5~kpc.
In  addition, calculated  distances  are in  good  agreement with  the
expansion parallaxes except LV Vul, LW Ser, and QV Vul.  Therefore, we
conclude that  our results are  in good agreement with  those obtained
from the expansion  parallax method, and the method  described in this
study  can be  used to  any  object as  long as  a reliable  and
  independent reddening estimate exists.

\begin{table*}
	\begin{center}
\caption{Results  for the  novae  which also  have expansion  parallax
  measurements. Here  $l$ and  $b$ are Galactic  coordinates, $E(B-V)$
  the reddening estimates obtained from the literature, $E(B-V)_\text{c}$ the
  reddening estimates  using the distances ($d_\text{exp}$)  from expansion
  parallaxes, $d_\text{c}$ indicates the distances calculated in this study.}
\label{table:exp}
	\begin{tabular}{lccccccl}
	\hline\hline
Object & $l$ & $b$ & $E(B-V)$ & $d_\text{c}$ & $d_\text{exp}$ & $E(B-V)_\text{c}$ & Ref$^a$ \\
 & ($\degrees$) & ($\degrees$) & (mag) & (kpc) &  (kpc) & (mag) &  \\
\hline
V1229 Aql & 40.537 & -5.437 &  $0.50\pm0.08$  &  $2.5\pm0.6$  & 2.4 & 0.48 &  \citet{Shafter97} \\
V1494 Aql & 40.974 & -4.742 &  $0.6\pm0.1$  &  $2.2\pm0.7$  & 1.2 & 0.36 &  \citet{Barsukovae13} \\
T Aur & 177.143 & 1.698 &  $0.42\pm0.08$  &  $1.08\pm0.37$  & 0.96 & 0.36 &  \citet{Slavine95} \\
V705 Cas & 113.66 & -4.096 &  $0.41\pm0.06$  &  $2.2\pm1.3$  & 2.5 & 0.42 &  \citet{Eyrese96} \\
V842 Cen & 316.574 & 2.453 &  $0.55\pm0.05$  &  $1.65\pm0.54$  & 1.2 & 0.5 &  \citet{DD00} \\
V1500 Cyg & 89.823 & -0.073 &  $0.45\pm0.07$  &  $2.0\pm0.6$  & 1.5 & 0.35 &  \citet{Slavine95} \\
v446 Her & 45.409 & 4.707 &  $0.37\pm0.04$  &  $1.5\pm0.47$  & 1.3 & 0.32 &  \citet{Cohen85} \\
CP Lac & 102.141 & -0.837 &  $0.27\pm0.06$  &  $1.67\pm0.61$  & 1.3 & 0.18 &  \citet{CohenR83} \\
BT Mon & 213.86 & -2.623 &  $0.24\pm0.06$  &  $1.84\pm0.47$  & 1.8 & 0.23 &  \citet{GillO98} \\
V959 Mon & 206.341 & 0.076 &  $0.38\pm0.06$  &  $2.3\pm0.6$  &  0.9-2.2  &  0.06-0.36  &  \citet{linfordetal15} \\
CP Pup & 252.926 & -0.835 &  $0.20\pm0.04$  &  $1.44\pm0.45$  & 1.14 & 0.12 &  \citet{DD00} \\
V1280 Sco & 351.331 & 6.553 &  $0.32\pm0.05$  &  $1.4\pm0.9$  & 1.6 & 0.38 &  \citet{Chesneaue08} \\
FH Ser & 32.909 & 5.786 &  $0.6\pm0.1$  &  $1.0\pm0.9$  & 0.95 & 0.56 &  \citet{GillO00} \\
LW Ser & 12.959 & 6.047 &  $0.39\pm0.1$  &  $1.27\pm0.40$  & 2.34 & 0.66 &  \citet{CohenR83} \\
V382 Vel & 284.167 & 5.771 &  $0.12\pm0.03$  &  $0.9\pm0.2$  & 0.8 & 0.11 &  \citet{Tomove15} \\
LV Vul & 63.303 & 0.846 &  $0.57\pm0.05$  &  $4.09\pm0.22$  & 2.5 & 0.32 &  \citet{Dellavalle98} \\
NQ Vul & 55.355 & 1.29 &  $0.92\pm0.2$  &  $2.0\pm0.5$  & 1.6 & 0.78 &  \citet{Slavine95} \\
QV Vul & 53.858 & 6.974 &  $0.4\pm0.05$  &  $1.2\pm0.3$  & 2.7 & 0.52 &  \citet{DD00} \\
\hline\hline
\end{tabular}
\footnotesize{
\begin{flushleft} $^a$ References for expansion parallaxes.
\end{flushleft} }
\end{center}
\end{table*}	

\begin{figure}
	\centering
	\includegraphics[width=0.46\textwidth]{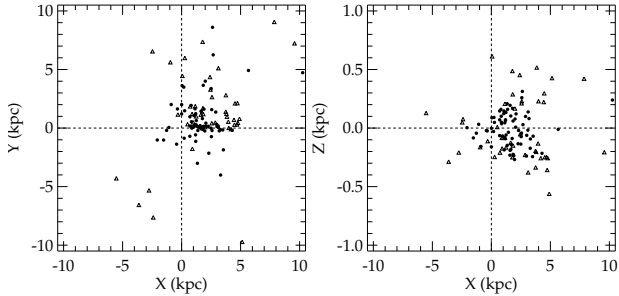}
	\caption{The      heliocentric       rectangular      Galactic
          distances. ($\bullet$) for which the distances are obtained,
          ($\vartriangle$) only lower limits calculated.}
	\label{Fig:1.7}
	\end {figure}

The  distances  of 55  Galactic  novae  which  do not  have  expansion
parallaxes are calculated (Table \ref{table:wexp}). We did not compare
our results with the distances compiled from the literature since they
generally   depend  on   statistical  relations   such  as   the  MMRD
relation. Distances based on  the MMRD relation \citep{DD00} have
  strong uncertainties  owing to the  scatter in the  relation itself,
  which is  about 0.6  mag with  deviations as large  as 1.6  mag.  In
  addition, the MMRD relation may not hold for our Galaxy as for other
  galaxies  \citep{Kasliwale11}.  Together  with these  55 novae the
total number of novae with accurately measured distances in our sample
increases to 73.  The distance  values range approximately from 0.7 to
11~kpc,  which  allows us  to  map  the  spatial distribution  of  the
Galactic novae.  The Sun  centred rectangular Galactic coordinates of
novae in our sample were also calculated and their projected positions
on the  Galactic plane ($X,Y$  plane) and  on plane perpendicular  to it
($X,Z$ plane)  are shown  in Fig.   \ref{Fig:1.7}.  From  their Galactic
distribution, we conclude that our method effectively works towards the
Galactic bulge  and up to  a $|z|<0.5$ kpc  in the Galactic  plane, as
excepted.

\begin{table*}
	\begin{center}
\caption{The distances of novae for which expansion parallax could not
  be  estimated.  The  column definitions  are  the same  as in  Table
  \ref{table:exp}.}
\label{table:wexp}
	\begin{tabular}{lcccc|lcccc}
	\hline\hline
Object & $l$ & $b$ & $E(B-V)$ & $d_\text{c}$ & Object & $l$ & $b$ & $E(B-V)$ & $d_\text{c}$  \\
 & ($\degrees$) & ($\degrees$) & (mag) & (kpc) &  & ($\degrees$) & ($\degrees$) & (mag) & (kpc) \\
\hline
CI Aql  & 31.688 & -0.812 &  $0.92\pm0.22$  &  $3.0\pm0.5$  &  V1187 Sco  & 355.724 & 1.451 &  $1.56\pm0.08$  &  $3.2\pm0.12$ \\
V1301 Aql  & 40.225 & -3.68 &  $0.61\pm0.05$  &  $2.2\pm0.3$  &  V1309 Sco  & 359.786 & -3.135 &  $0.55\pm0.05$  &  $2.5\pm0.4$\\
V1493 Aql  & 45.908 & 2.155 &  $0.57\pm0.14$  &  $1.72\pm0.66$  &  V1324 Sco  & 357.425 & -2.872 &  $1.16\pm0.12$  &  $4.3\pm0.9$ \\
V1721 Aql  & 40.972 & -0.083 &  $3\pm0.6$  &  $7.5\pm2.0$  &  V745 Sco  & 357.358 & -3.999 &  $0.84\pm0.15$  &  $3.50\pm0.85$ \\
OY Ara  & 333.902 & -3.937 &  $0.32\pm0.06$  &  $1.32\pm0.7$  &  V977 Sco  & 357.603 & -2.998 &  $0.92\pm0.1$  &  $3.3\pm0.6$ \\
V1065 Cen  & 293.984 & 3.613 &  $0.47\pm0.05$  &  $3.3\pm0.5$  &  V992 Sco  & 343.824 & -1.602 &  $1.3\pm0.1$  &  $2.63\pm0.35$ \\
V868 Cen  & 309.508 & -1.028 &  $1.72\pm0.1$  &  $5.2\pm0.4$  &  V368 Sct  & 24.669 & -2.629 &  $0.5\pm0.1$  &  $1.94\pm0.95$ \\
AR Cir  & 317.039 & -0.372 &  $2.0\pm0.4$  &  $3.15\pm0.56$  &  V443 Sct  & 27.218 & -2.421 &  $0.40\pm0.05$  &  $1.4\pm0.5$ \\
BY Cir  & 315.022 & -3.723 &  $0.13\pm0.06$  &  $0.99\pm0.22$  &  V476 Sct  & 24.741 & 1.213 &  $1.9\pm0.1$  &  $11.3\pm1.9$ \\
V394 CrA  & 352.822 & -7.723 &  $0.2$  &  :2.0  &  V496 Sct  & 25.284 & 1.768 &  $0.57\pm0.07$  &  $3.2\pm0.8$ \\
Q Cyg  & 89.929 & -7.552 &  $0.26\pm0.06$  &  $1.22\pm0.9$  &  MU Ser  & 14.066 & 5.555 &  $0.4\pm0.1$  &  $0.73\pm0.53$\\
V2274 Cyg  & 73.041 & 1.991 &  $1.33\pm0.1$  &  $9.0\pm2.0$  &  WY Sge  & 53.368 & -0.739 &  $1.6\pm0.3$  &  $4.2\pm0.4$ \\
V2275 Cyg  & 89.317 & 1.391 &  $1.0\pm0.1$  &  $3.62\pm0.47$  &  V1172 Sgr  & 7.639 & 3.336 &  $0.4\pm0.08$  &  $0.86\pm0.14$\\
V2467 Cyg  & 80.069 & 1.842 &  $1.4\pm0.2$  &  $1.5\pm0.3$  &  V4077 Sgr  & 7.427 & -8.319 &  $0.32\pm0.03$  &  $1.6\pm1$\\
V2468 Cyg  & 66.808 & 0.245 &  $0.78\pm0.1$  &  $6.8\pm1.0$  &  V4160 Sgr  & 0.2 & -6.968 &  $0.35\pm0.04$  &  $1.85\pm1.2$\\
V2491 Cyg  & 67.229 & 4.353 &  $0.23\pm0.01$  &  $2.1\pm1.4$  &  V4169 Sgr  & 4.558 & -6.964 &  $0.35\pm0.05$  &  $1.9\pm0.7$\\
V407 Cyg  & 86.983 & -0.482 &  $0.55\pm0.05$  &  $3.5\pm0.3$  &  V4332 Sgr  & 13.628 & -9.4 &  $0.32\pm0.1$  &  $1.14\pm1.0$\\
IM Nor & 327.098 & 2.485 &  $0.8\pm0.2$  &  $2.4\pm0.8$  &  V4444 Sgr  & 3.817 & -3.388 &  $0.45\pm0.1$  &  $2.27\pm0.58$\\
V382 Nor  & 332.29 & -0.995 &  $0.85\pm0.15$  &  $4.0\pm0.53$  &  V4633 Sgr  & 5.128 & -6.231 &  $0.26\pm0.05$  &  $1.36\pm0.46$\\
V2214 Oph  & 355.372 & 5.724 &  $0.73\pm0.1$  &  $2.6\pm1.0$  & V4643 Sgr  & 3.346 & -0.337 &  $1.47\pm0.2$  &  $3.1\pm0.2$ \\
V2295 Oph  & 2.378 & 6.973 &  $0.75\pm0.2$  &  $1.6\pm1.3$  &  V5113 Sgr  & 3.721 & -4.091 &  $0.1\pm0.02$  &  $0.95\pm0.21$\\
V2487 Oph  & 6.604 & 7.775 &  $0.38\pm0.08$  &  $1.08\pm0.28$  &  V5115 Sgr  & 6.046 & -4.567 &  $0.53\pm0.05$  &  $3\pm1$\\
V2576 Oph  & 356.2 & 5.369 &  $0.62\pm0.12$  &  $1.8\pm1$  &  V5116 Sgr  & 2.136 & -6.833 &  $0.23\pm0.06$  &  $1.6\pm0.8$\\
V2615 Oph  & 4.148 & 3.302 &  $0.95\pm0.15$  &  $2.1\pm0.8$  &  V5117 Sgr  & 3.337 & -1.429 &  $0.5\pm0.1$  &  $1.45\pm0.33$ \\
V2672 Oph  & 1.019 & 2.529 &  $1.6\pm0.1$  &  $3.12\pm0.69$  &  V5558 Sgr  & 11.611 & 0.207 &  $0.8\pm0.1$  &  $2.1\pm0.4$ \\
V2674 Oph  & 357.842 & 3.579 &  $0.7\pm0.1$  &  $1.65\pm0.38$  & V732 Sgr  & 2.529 & -1.188 &  $0.81\pm0.16$  &  $3.0\pm0.2$ \\
V972 Oph  & 359.376 & 2.428 &  $0.5\pm0.1$  &  $0.87\pm0.15$  &  CK Vul  & 63.383 & 0.99 &  $0.75\pm0.05$  &  $4.48\pm0.24$ \\
V529 Ori  & 188.935 & -1.937 &  $0.32\pm0.02$  &  $1.29\pm0.35$ &  &  &  &  &  \\
\hline\hline
\end{tabular}
\end{center}
\end{table*}

The reddening estimates of some  novae exceed the lower/upper limit of
the reddening curve derived from 2MASS, UKIDSS or VISTA photometry. In
such  cases, we  conclude  that the  distance of  the  nova should  be
lower/higher  than  the  calculated   distance  corresponding  to  the
lower/upper limit of the  reddening-distance relation. The lower/upper
limits   of   the   calculated   distances   are   listed   in   Table
\ref{table:lowliml}.  From Fig.  \ref{fig:fig1.1}, it can be seen that
the  galactic  latitudes of  these  novae  are generally  higher  than
$|b|\ge2\degrees$, or  they are nearby.  In both cases,  extinction to
the novae is low, making it  difficult to produce an estimate from the
RC photometry.  Lower limits  of the  distances indicate  that
these novae  are generally  more than 0.2~kpc  away from  the Galactic
plane (Fig. \ref{Fig:1.7}).

\begin{table*}
	\begin{center}
\caption{Results of  novae for  which only  lower/upper limits  of the
  distance could be calculated. The column definitions are the same as
  in Table \ref{table:exp}.}
\label{table:lowliml}
	\begin{tabular}{lccccl|lccccl}
	\hline\hline
Name & $l$ & $b$ & $E(B-V)$ & $d_\text{c}$ & ref$^a$ & Name & $l$ & $b$ & $E(B-V)^a$ & $d_\text{c}$ & Ref$^a$ \\
 & ($\degrees$) & ($\degrees$) & (mag) & (kpc) &  & &($\degrees$) & ($\degrees$) & (mag) & (kpc) & \\\hline
V1370 Aql & 38.813 & -5.946 &  $0.35\pm0.05$  &  $\geq$1.6 & 1	& V597 Pup & 252.529 & 0.546 & 0.3 &  $\geq$8.0  & 21 \\
V1419 Aql$^c$ & 36.811 & -4.1 &  0.7-1.2  &  $\geq$4.70  & 2 	& V960 Sco & 358.72 & -3.506 & 1.3 &  $\geq$4.2  & 22 \\
V1425 Aql & 33.011 & -3.893 &  0.65,0.76  &  $\geq$2.20  & 3 	& V373 Sct & 26.504 & -4.397 &  $0.32\pm0.05$  &  $\geq$1.1  & 23 \\
V1680 Aql & 45.779 & 3.56 & 1 &  $\geq$3.70  & 4  	& V444 Sct & 25.012 & -2.854 &  $0.79-1.05$   &  $\geq$ 5.0  & 24 \\
V1722 Aql & 49.085 & 2.014 &  $1.35\pm0.1$  &  $\geq$12.0  & 5 	& V463 Sct & 17.838 & -2.917 &  $0.8\pm0.1$  &  $\geq$4.0  & 25 \\
V1830 Aql & 37.097 & -0.985 & 2.6 &  $\geq$12.0  & 6 	& V475 Sct & 24.202 & -3.947 &  $0.69\pm0.05$  &  $\geq$5.20  & 26 \\
V603 Aql & 33.164 & 0.829 &  $0.08\pm0.02$  &  $\leq$0.60  & 7 	& V477 Sct & 20.568 & -2.789 &  $1.2\pm0.2$  &  $\geq$4.0  & 27 \\
IV Cep & 99.614 & -1.638 &  $0.65\pm0.05$  &  $\geq$5.70  & 1 	& RT Ser & 13.895 & 9.971 &  $0.64\pm0.1$  &  $\geq$1.2  & 28 \\
V809 Cep & 110.647 & 0.399 &  $1.7\pm0.34$  &  $\geq$7.0  & 6	 & V378 Ser & 14.167 & 7.428 & 0.74 &  $\geq$4.00  & 29 \\
CP Cru$^b$ & 297.872 & 2.223 &  $1.9\pm0.1$  &  $\geq$11.0  & 8 	& HM Sge$^b$ & 53.567 & -3.15 &  0.43-0.75  &  $\geq$1.5  & 30 \\
V1974 Cyg$^b$ & 89.134 & 7.819 &  $0.26\pm0.03$  &  $\geq$4.50  & 9 	& V3888 Sgr$^b$ & 9.083 & 4.66 &  $1.0\pm0.2$  &  $\geq$2.6  & 31 \\
V2361 Cyg & 76.406 & 3.676 & 1.2 &  $\geq$7.60  & 10 	& V3890 Sgr & 9.204 & -6.443 &  0.8-1  &  $\geq$5.00  & 32 \\
V2362 Cyg & 87.372 & -2.357 &  $0.56\pm0.05$  &  $\geq$6.0  & 11 	& V4157 Sgr & 5.321 & -3.066 &  0.94-1.07  &  $\geq$4.8  & 24 \\
V838 Her & 43.316 & 6.619 &  0.4-0.6  &  $\geq$2.50  & 12 	& V4171 Sgr & 9.38 & -4.54 &  $0.61\pm0.12$  &  $\geq$2.90  & 24 \\
DK Lac$^b$ & 105.237 & -5.351 &  $0.22\pm0.06$  &  $\geq$1.20  & 7 	& V4642 Sgr & 8.991 & 2.83 &  $1.51\pm0.06$  &  $\geq$4.5  & 33 \\
V838 Mon & 217.798 & 1.052 & 0.87 &  $\geq$7.0  & 13 	& V4739 Sgr & 3.211 & -7.966 &  $0.46\pm0.04$  &  $\geq$1.80  & 34 \\
GQ Mus & 297.212 & -4.996 &  $0.45\pm0.05$  &  $\geq$2.0  & 14 	& V5579 Sgr & 3.734 & -3.023 & 1.2 &  $\geq$4.7  & 35 \\
V2575 Oph & 2.415 & 4.782 &  $1.42\pm0.28$  &  $\geq$2.0  & 15 	& V5588 Sgr & 7.84 & -1.884 &  $1.56\pm0.1$  &  $\geq$4.0  & 36 \\
V2670 Oph & 3.666 & 3.78 & 1.3 &  $\geq$4.50  & 16 	& PW Vul$^b$ & 61.098 & 5.197 &  $0.57\pm0.03$  &  $\geq$5.00  & 37 \\
V2671 Oph & 0.199 & 3.285 & 2 &  $\geq$4.0  & 17 	& QU Vul$^b$ & 68.511 & -6.026 &  $0.55\pm0.05$  &  $\geq$2.00  & 14, 38 \\
V351 Pup & 53.368 & -0.739 &  $0.72\pm0.01$  &  $\geq$4.1  & 18 	& V458 Vul & 58.633 & -3.617 &  0.50-0.63  &  $\geq$6.00  & 1,39 \\
V445 Pup$^b$ & 241.124 & -2.192 &  $0.51\pm0.1$  &  $\geq$7.5  & 19 	& V459 Vul & 58.214 & -2.167 &  $0.86\pm0.12$  &  $\geq$2.2  & 40 \\
V574 Pup & 242.57 & -1.993 &  $0.5\pm0.1$  &  $\geq$6.0  & 20 	& NSV 11749 & 34.872 & -3.556 &  $0.75\pm0.15$  &  $\geq$2.0  & 41 \\
\hline\hline
\end{tabular}
\footnotesize{ 
\begin{flushleft} $^{a}$ References for reddening estimates:  1 -- \citet{HachisuK16};
2 -- \citet{Lynche95};
3 -- \citet{Arkhipovae02,Kamathe97};
4 -- \citet{LloydG02};
5 -- \citet{Munarietal10};
6 -- \citet{Munarie14};
7 -- \citet{SelvelliG13};
8 -- \citet{Lyke03};
9 -- \citet{Chochole97b};
10 -- \citet{Russelle05};
11 -- \citet{Sivieroe06,Munarie08b,Mazuke06,Lynche08b,Russelle06};
12 -- \citet{VanlandinghamE96};
13 -- \citet{Munariel05,Tylenda05};
14 -- \citet{Hachisu14};
15 -- \citet{Ruselle06};
16 -- \citet{Russelle08c};
17 -- \citet{Rudye08b};
18 -- \citet{Saizare96};
19 -- \citet{IijimaN08};
20 -- \citet{Nesse07,Schwarze11};
21 -- \citet{Nesse08b};
22 -- \citet{RichtlerL86};
23 -- \citet{VanGenderen77,Warner87};
24 -- \citet{Williams94};
25 -- \citet{Katoe02};
26 -- \citet{Chochole05};
27 -- \citet{Mazuke05};
28 -- \citet{Rudye99};
29 -- \citet{Russelle05b};
30 -- \citet{SchmidS90};
31 -- \citet{VogtM77};
32 -- \citet{Gonzalezr92,AnupamaS94,Schaefer10};
33 -- \citet{Venturinie04};
34 -- \citet{Living01};
35 -- \citet{Russelle08b};
36 -- \citet{Muntarie15};
37 -- \citet{Andrea91,Williamse96,Burnashev95};
38 -- \citet{Rosinoe92,Saizare92,DellaValle97};
39 -- \citet{Lynche07,Wessone08,Poggiani08};
40 -- \citet{Poggiani10};
41 -- \citet{BondK12}
\\
$^b$ Calculated reddening for given expansion parallaxes: 
Cp Cru: $E(B-V)_\text{c}=0.5$ mag for 3.2 kpc \citep{DD00};
V1974 Cyg: $E(B-V)_\text{c}=0.2$ mag for 1.8 kpc	\citep{Chochole97b};
DK Lac: $E(B-V)_\text{c}=0.21$ mag for 3.9 kpc \citet{Cohen85}; 
V445 Pup: $E(B-V)_\text{c}=0.48$ mag for 8.2 kpc \citep{Woudte09}; 
HM Sge: $E(B-V)_\text{c}=0.44$ mag for 1.78 kpc \citep{HackP93};
V3888 Sgr: $E(B-V)_\text{c}=0.82$ mag for 2.5 kpc \citep{DD00}; 
PW Vul: $E(B-V)_\text{c}=0.34$ mag for 1.8 kpc \citep{DD00}; 
QU Vul: $E(B-V)_\text{c}=0.46$ mag for 1.75 kpc \citep{DD00}.\\
$^{c}$ With using $E(B-V)=0.5$ \citep{Hachisu14}, $d_\text{c}\geq2.5$ kpc.
\end{flushleft} }
\end{center}
\end{table*}

\section{CONCLUSIONS}
\label{concsec}
In this paper, we measured distances  of a large number of novae
utilizing  a method  that  relies on  the Galactic  reddening-distance
relation for  the line-of-sight  of a given  nova and  its independent
reddening measurement. The specific reddening-distance relation in the
line of sight of a nova was derived using the unique properties of the
RC  stars.  Especially  for directions  within  the thin  disc of  our
Galaxy, this method  proves to be very effective that  makes it
easily applicable  to a  sample of sources for which distances
could not  be found using  other methods, and  it depends on  only one
parameter:  the  reddening.   Using  this  method,  we  were  able  to
determine the distances of 73 Galactic  novae, and set lower limits on
the distances of  another 46 systems.  Although only  a limited sample
is available, we compared our distance estimates with those found from
the  expansion parallaxes  and concluded  that the  distances obtained
from the two methods are in agreement with each other within errors.

The  Galactic  novae in  our  study  are  mainly located  towards  the
Galactic  bulge  and concentrate  close  to  the Galactic  plane  with
typical vertical  distances of  $|z|\le0.3$ kpc. This  distribution is
consistent  with  the  Galactic  structure analyses  of  novae,  which
resulted  in  a  scale   height  of  $\sim125$  pc  \citep{Duerbeck81,
  DellaVella92, Ozdonmez15}.  Nevertheless,  the galactic distribution
of novae  in our study could  simply be a result  of the observational
selection  effects,  since  our   distance  measurement  method  works
effectively only  for the novae  within the Galactic  disc, especially
towards the bulge.

In the  distance calculations presented here,  the largest uncertainty
arises from the uncertainties in the reddening estimates.  Thus,
we considered  all the reddening  estimates for a particular  nova and
compared them.  Some  novae have only one or a  rather crude reddening
estimate, which may  not represent the actual value  owing to blending
effect, contamination  of extra reddening  from dust in  the expansion
shell,  low  resolution  observations, underestimated  coefficient  of
calibration  between  any  spectral  line  or  colour.   For  example,
reddening estimates of V5558~Sgr, BY~Cir and V2295~Oph (see Appendix A
for details) vary in a large range, because different extinction
measurement  methods  result  in   different  values,  which  lead  to
different distances.  For these  systems, the distances estimated from
the distance-reddening  relations in this  study can not  be reliable.
Nevertheless,  for novae  which  have  more than  one  reddening
measurement, standard  deviation of  the reddening  differences from
the mean  was calculated  as 0.20 mag,  which is  roughly consistent
with the mean  value of the reddening  errors ($\sim0.12$).  Future
spectroscopic observations  of a large  sample of novae are  needed to
decrease the uncertainty of reddening estimates.

Determination of distances of Galactic novae in such a systematic way 
has several important implications on understanding the spatial distribution, 
luminosity function of these systems. For example, 
as mentioned earlier, the existence of an MMRD relation for the 
Galaxy has long been questioned. Using the Galactic nova sample with distances 
calculated in this study and parameters of the outburst light curves, the MMRD relation 
of Galactic novae can be examined. Such a study is out of the scope of this paper but will be presented elsewhere.
In addition,  the distance-reddening  maps derived  for 119  novae can
also be  used to estimate reddening  or distance for any  source whose
location is close to the position of a nova in our sample.

\section*{Acknowledgements}

We  acknowledge the  anonymous referee  for helpful  comments on  this
paper. AO  wishes to thank  to the members of the Instituto de
Astrof\'isica de Canarias for their hospitality during his visit to La
Laguna, Tenerife.   We thank Dimitrios  Psaltis and Feryal  \"Ozel for
their    support   on    the    calculation    of   the    probability
distributions. This work has been  supported in part by the Scientific
and Technological Research Council of Turkey (T\"UB\.ITAK) 114F271. TG
was  supported by  Scientific  Research Project  Coordination Unit  of
Istanbul University, Project numbers 49429 and 57321. This publication
makes use of  data products from the Two Micron  All Sky Survey, which
is a joint project of the University of Massachusetts and the Infrared
Processing  and Analysis  Center/California  Institute of  Technology,
funded by  the National Aeronautics  and Space Administration  and the
National Science Foundation. This work is based on data from the UKIRT
Infrared Deep Sky Survey,  UKIDSS (www.ukidss.org). We also gratefully
acknowledge  use  of  data  from  the ESO  Public  Survey  program  ID
179.B-2002  taken with  the VISTA  telescope, data  products from  the
Cambridge Astronomical Survey Unit. This  research has made use of the
SIMBAD, and NASA's Astrophysics Data System Bibliographic Services.

\appendix
\section{Individual Objects}
\label{appAsec}
This section  presents the  summary and  discussions of  earlier
interstellar reddening  and distance  measurements for each  nova 
  used in this study.

\subsection{Sources with Expansion Parallax Measurements}
\subsubsection{V1229 Aql}
The  distance and  reddening of  V1229~Aql have  been well-determined.
\citet{DellaValle93}  adopted   an  average  value  of   reddening  as
$E(B-V)=0.50\pm0.08$~mag    from    various    reddening    estimates:
$E(B-V)=0.49\pm0.05$~mag via a  comparison of observed colour of
nova    with    intrinsic     colours,    $E(B-V)=0.79$    mag    from
H$_\alpha$/H$_\beta$,       $E(B-V)=0.36$      mag       from      He~\textsc{i}
$\lambda$5876/$\lambda$4471,       $E(B-V)=0.25$        mag       from
$\lambda$6678/$\lambda$4471,  $E(B-V)=0.65$ and  $0.30$ mag  both from
the    equivalent     widths    of    Na~\textsc{i}     lines.     Furthermore,
\citet{DellaValle93}   determined  the   distance   of  V1229~Aql   as
$d=2.1\pm0.9$~kpc from  the shell parallax  method. \citet{Duerbeck81}
obtained the  reddening of  the nova as  $E(B-V)=0.52\pm0.13$~mag from
the interstellar line strengths, and  the distance as $d=1.73$~pc from
expansion   velocities   based   on  the   Ca~\textsc{ii}~K   line.    Finally,
$E(B-V)=0.59\pm0.02$~mag was  given by  \citet{Miroshnichenko88} based
on intrinsic colour of the nova during the stabilization stage.  Since
the reddening  evolution is nearly  constant at $\sim0.6$~mag  for the
distances higher  than 4~kpc,  we can  only set a  lower limit  on the
distance of  V1229~Aql for the  reddening estimates larger than
0.6~mag.   In this  study,  we  adopted the  $E(B-V)=0.50\pm0.08$~mag,
given by \citet{DellaValle93}, which is  in a good agreement with that
reported  by  \citet{Duerbeck81} and  \citet{Miroshnichenko88}.  Using
this value, we calculated the distance of V1229 Aql as $d_\text{c}=2.5\pm0.6$
kpc, which is  consistent with the 2.4~kpc  given by \citet{Shafter97}
from expansion velocities in \citet{CohenR83,Cohen85}, and $2.1\pm0.9$
kpc obtained by \citet{DellaValle93}, within the uncertainties of
  the measurements.
\subsubsection{V1494 Aql}
\citet{IijimaE03}  used  the  equivalent widths  of  the  interstellar
absorption  components  of   Na~\textsc{i}  D1  and  D2,   and  estimated  the
interstellar reddening  towards V1494~Aql as  $E(B-V)=0.6\pm0.1$ mag,
which  is consistent  with 0.66~mag obtained  by \citet{Arkhipovae02}
from spectral line ratios. We used the reddening as $E(B-V)=0.6\pm0.1$
mag and than  calculated the distance of V1494~Aql as $d_\text{c}=2.2\pm0.7$
kpc.   This  result   is  larger   than  $1.2\pm0.2$   kpc  given   by
\citet{Barsukovae13}  from   the  expansion  velocity   of  $\sim0.24$
arcsec/year  which depends  on  only one  structure around  H$_\alpha$
emission line.
\subsubsection{T Aur}
The  reddening  of T~Aur  has  been  reported by  many  authors
\citep{Gallaghere80,  Duerbeck81,Warner87,Gilmozzie94, BE94,  DiazB97,
  Shafter97,     DD00,     Selvelli04,    Pueblae07,     SelvelliG13}.
\citet{SelvelliG13} recently calculated the reddening as $E(B-V)
= 0.42\pm0.08$~mag, using 2200 {\AA} feature.  Using the distance of T
Aur  as  $d=0.96\pm0.22$ kpc  from  the  expansion parallax  given  by
\citet{Slavine95}, we  also obtained the reddening  as $E(B-V)_\text{c}=0.36$
mag,  which  is consistent  with  that  given by  \citet{SelvelliG13},
within errors.   We used $E(B-V)  = 0.42\pm0.08$ mag to  calculate the
distance of  T Aur as  $d_\text{c}=1.08\pm0.37$ kpc, in  agreement with
$960\pm220$ pc by \citep{Slavine95}, as well.
\subsubsection{V705 Cas}
The reddening towards  V705~Cas was estimated by  \citet{Hrice98} to be
$E(B-V)= 0.38$  and 0.43~mag, from  an inter-comparison of  the colour
indices of the stars surrounding  the nova and using intrinsic colours
at    the   stabilization    stage   \citep{Miroshnichenko88}, 
respectively.   However  \citet{Hachisu14}  mentioned  that  colour
excesses  at  maximum  and $t_2$  \citep[data  taken  from][]{Hrice98}
indicate $E(B-V)=0.33$ and 0.36  mag, and adopted $E(B-V)=0.45\pm0.05$
mag  calculated from  colour-colour  evolution of  the  nova in  their
study.  \citet{Hauschildte94}  assumed that  the total (optical  + UV)
luminosity during the initial outburst  is constant, and than obtained
$E(B-V)=0.5$ mag.   On the other hand,  \citet{Arkhipovae99} used
the  Balmer decrement  and  statistical relations  to estimate  the
reddening as $E(B-V)=0.98$ and $0.7$ mag, respectively.  We
did  not   take  into   account  the   reddening  estimate   given  by
\citet{Arkhipovae99},  since  they   concluded  that  the  discrepancy
between  these two  values could  result from  the reddening  owing to
circumstellar dust shell and  their values systematically exceed
other  estimates.  Another  adopted reddening  value for  V705~Cas was
reported as 0.67~mag by \citet{Elkin95}, who used the colour excess at
maximum.   The  information  about  reddening  estimates  given  above
indicate  two  mean  reddening  values  $E(B-V)=0.41\pm0.06$  and
$0.84\pm0.14$ mag.   By using these mean  values, we calculated
the distance  to be $d_\text{c}=2.2\pm1.3$ and  $d_\text{c}\geq5$ kpc, respectively.
Note that the reddening evolution  after 2~kpc becomes shallow leading
the upper limit of the distance to be uncertain.  The smaller distance
calculated here  is much more  consistent with distances  from earlier
expansion  parallax measurements  as 2.9  \citep{Diaze01} and  2.5~kpc
\citep{Eyrese96}.  Thus we conclude that the reddening and distance of
V705~Cas   are   the    most   likely   $E(B-V)=0.41\pm0.06$~mag   and
$d_\text{c}=2.2\pm1.3$~kpc, respectively.
\subsubsection{V842 Cen}
The mean interstellar reddening towards the moderately fast nova
V842~Cen     was     given    as     $E(B-V)=0.55\pm0.05$~mag     from
H$_\alpha\text{/H}_\beta$  ratio \citep{deFreitas89},  from equivalent
width of the NaI D lines  \citep{Sekiguchie89} and from the 2200 {\AA}
feature \citep{Andrea91}. Using this reddening estimate, we calculated
the  distance  of  V482~Cen  as $d_\text{c}=1.65\pm0.54$  kpc,  which  is  in
agreement   with   $1.2\pm0.1   \    \text{and}   \   1.3\pm0.5$   kpc
\citep{DD00,GillO98}  obtained from  expansion parallaxes,  within the
uncertainties.
\subsubsection{V1500 Cyg}
V1500~Cyg is a  well-known and extremely fast  nova.  The interstellar
reddening towards V1500~Cyg has been calculated in a number of studies;
\citet{Tomkin76}   derived  $E(B-V)   =  0.45   \pm  0.07$   mag  from
interstellar KI  7699 {\AA}  line while \citet{McLean76}  used optical
polarization to  determine $E(B-V) = 0.45  \pm 0.02$ mag which  is the
same as found by \citet{Hachisu14} from colour-colour evolution. These
values are in  agreement with that estimated  by \citet{Ferland77} who
compared observational and theoretical flux  ratio to derive the
colour excess $E(B-V) = 0.5 \pm$ 0.05 mag.  \citet{Wue77} calculated a
reddening of $0.69 \pm 0.03$~mag from the 2200~{\AA} feature. Although
estimation of the reddening from the 2200~{\AA} feature gives reliable
results, this feature could be  affected by line emission as mentioned
in \citet{Ferland77}. Thus, it can  be concluded that the reddening is
less than 0.57 mag. Therefore we adopted the reddening to be $E(B-V) =
0.45 \pm 0.07$ mag given by \citet{Tomkin76}, which is consistent with
that given above  except the value obtained  from the 2200~{\AA}
feature.  Using the reddening-distance relation obtained in this
study, we  calculated the  distance of  V1500~Cyg as  $2.0 \pm0.6$~kpc
which is consistent with those found by \citet[1.5 kpc]{Slavine95} and
\citet[1.56 and 2.11 kpc]{Wadee91} within errors.
\subsubsection{V446 Her}
The     reddening    towards     V446~Her     was    determined     as
$E(B-V)=0.46\pm0.04$~mag  from the  value at  the stabilization  stage
\citep{Miroshnichenko88},    $0.40\pm0.05$~mag   from    colour-colour
evolution   \citep{Hachisu14},   0.35  \citep{Gilmozzie94},   0.25
\citep{Selvelli04} and $0.38\pm0.04$~mag  \citep{SelvelliG13} from the
2200  {\AA} feature.   We  adopted   the  mean  of  these  estimates,
$0.37\pm0.04$ mag, as the reddening of V446 Her. This mean value is in
agreement with  the results of  \citet{SelvelliG13}, which
depends  on  more reliable  methods.  Using  this mean  reddening,  we
derived  a  distance  of $d_\text{c}=1.50\pm0.47$  kpc,  which  is  in
agreement with the distance calculated  from the expansion parallax by
\citet{Cohen85}, $d=1.3$ kpc, within errors.
\subsubsection{CP Lac}
The   interstellar    reddening   of    CP~Lac   was    estimated   as
$E(B-V)=0.26$~mag  from  Na~\textsc{i}  lines  \citep{CohenR83}  and  as
$0.28\pm0.06$    mag   from  the 2200   {\AA}    feature
\citep{SelvelliG13}.  We  used the  mean value  of these  estimates as
$E(B-V)=0.27\pm0.06$ mag and  calculated the distance of CP  Lac to be
$d_\text{c}=1.67\pm0.61$ kpc, which is  in agreement with  $d=1.3$ kpc
estimated from expansion parallax \citep{CohenR83}
\subsubsection{BT Mon}
Using the 2200 {\AA} feature, \citet{SelvelliG13} calculated the
colour  excess  as $E(B-V)=0.24\pm0.06$~mag.   Since  their  UV
spectral analysis for 18 old novae supersede the previous ones derived
in   earlier   studies  \citep{Gilmozzie94,Selvelli04},   we   adopted
$E(B-V)=0.24\pm0.06$  mag   \citep{SelvelliG13}  and   calculated  the
distance  of  BT  Mon  as  $d_\text{c}=1.84\pm0.47$ kpc,  which  is  in  good
agreement  with  $d=1.8$  kpc  estimated  using  expansion  parallaxes
\citep{Marshetal83,GillO98}.
\subsubsection{V959 Mon}
\Citet{Munarietal2013} investigated  the expansion  evolution of
this new nova, Nova~Mon~2012, and estimated the interstellar reddening
as $E(B-V)=0.38\pm0.01$ mag from the equivalent width of the Na~\textsc{i}
line. A  preliminary estimate was given as  $E(B-V)=0.30$ mag by
\citet{Munarietal12}.    This   estimate   is  smaller  than
$E(B-V)=0.8\pm0.05$  mag  obtained  from the  hydrogen  column
density \citep{Shoreea13},  which indicates a  distance of
$d_\text{c}\geq6$ kpc.  Using $E(B-V)=0.38\pm0.01$  mag, the distance of V959
Mon   was   calculated  as   $d_\text{c}=2.3\pm0.6$   kpc   in  this   study.
\citet{linfordetal15} investigated  the ejecta  of V959 Mon  to obtain
the  expansion  parallax.   In their  study,  non-spherical  expansion
implies a distance between $0.9\pm0.2$ and $2.2\pm0.4$ kpc. They noted
that the  most probable  distance is $1.4\pm0.4$  kpc.  Our  result is
consistent with upper limit given by \citet{linfordetal15}.
\subsubsection{CP Pup}
The interstellar reddening  towards CP~Pup has been reported  to be in
the    0.20$-$0.26~mag    range    by     a    number    of    authors
\citep{Duerbeck81,Gilmozzie94,BE94,DiazB97,DD00,Selvelli04,Pueblae07,
  SelvelliG13}  based  on  galactic  extinction  maps  and  the
2200~{\AA}    feature.    Among    these   studies,    the   one    by
\citet{SelvelliG13}  is especially  important  since their  new
measurements are based on the 2200 {\AA} feature, and 
  they indicate  a reddening of $E(B-V)=0.20\pm0.04$  mag.  From this
value, we  calculated the distance  of CP~Pup to  be $d_\text{c}=1.44\pm0.45$
kpc  using  a  field with  a  radius  of  1.2  ${\deg}$ on  the  2MASS
CMD. Despite the  fact that  we had to  use a  relatively larger
  field of view,  our result is consistent with the  $1.14$ kpc given
by  the  last  expansion  parallax study  \citep{DD00},  and  previous
distance estimates \citep{Duerbeck81,Williams82,  GillO98} that placed
CP Pup  at distances of  1.5, 1.6, and $1.8\pm0.4$  kpc, respectively.
Since,  \citet{CohenR83}  neglected  the shell  geometry,  which
  resulted in a smaller expansion  velocity, the distance estimate of
0.85 kpc in their study is smaller than ours and others.  So, we
conclude that our result is reliable.
\subsubsection{V1280 Sco}
\citet{Chesneaue08} assumed that the secondary star of V1280 Sco
have a temperature of  $T=7000$ K and colour  as $B-V\sim$0/0.3
mag (for an A or F spectral type star), and estimated the reddening as
$E(B-V)=0.3$  mag from  the difference  between observed  and spectral
colours.  This estimate is consistent  with $E(B-V)=0.35$ mag given by
\citet{HachisuK16} using the intrinsic  colour of V1280 Sco.  However,
      \citet{Puetteretal07} calculated the reddening of V1280~Sco
        as $E(B-V)=1.7$  mag from O~\textsc{i}  lines, but they noted  that the
        estimated reddening is  due in part to the  dust shell. 
        This  measurement indicates  that  the distance  of V1280  Sco
        should be  larger than  5 kpc according  to reddening-distance
        relation obtained  in this study. In  addition, we estimated
      the reddening  as 0.25  and 0.38  mag for  the distances  of 1.1
      \citep{Naitoe12} and  1.6 kpc \citep{Chesneaue08}  obtained from
      expansion  parallaxes.  Therefore,  we  used  mean reddening  as
      $E(B-V)=0.32\pm0.05$ mag,  and calculated the distance  of V1280
      Sco as $d_\text{c}=1.4\pm0.9$ kpc, which  is in good agreement with the
      estimates  given  by  \citet{Naitoe12}  and  \citet{Chesneaue08}
      within uncertainties.
\subsubsection{FH Ser}
\citet{DellaValle97} investigated the reddening towards FH~Ser through
three  methods: $E(B-V)=0.82$  mag using  the  colour of  the nova  at
maximum, 0.61 mag from the line ratio of H$_\alpha$/H$_\beta$, 0.5 mag
from the equivalent width of Na~\textsc{i} $\lambda5890$.  \citet{DellaValle97}
adopted  an averaged  value of  $E(B-V)=0.64\pm0.16$ mag  which is  in
agreement,    within   errors,    with   both    the   0.6    mag   by
\citet{Kodaira70}, which  was  obtained  from the  interstellar
reddening  relation   and  0.74  mag  adopted   by  \citet{Weighte94}.
\citet{Hachisu14}  adopted a  similar reddening  for FH~Ser  as
$E(B-V)=0.6$~mag obtained from the intrinsic colour evolution of
this    nova.     In   addition,    \citet{HutchingsF93}    calculated
$E(B-V)\sim0.9$ mag  obtained from  the diffuse interstellar  lines at
$\lambda$5780 and  $\lambda$5796 in the early  post-maximum of FH~Ser,
but  this  value  is  higher   than  others.   We  derived  the
reddening-distance  relation  and   obtained  a distance  of
$d_\text{c}=1.0\pm0.9$ kpc, using $E(B-V)=0.6\pm0.1$ mag, which is consistent
with  the spectroscopic  reddening  measurement  and intrinsic  colour
evolution  given above.   Our result  is in  very good  agreement with
$950\pm50$ pc  \citep{GillO00} obtained from new  HST and ground-based
observations, and  also consistent  with other measurements  that
  were   obtained   from   expansion   parallaxes;   $920\pm130$   pc
\citep{Slavine95}, $870\pm90$  pc \citep{DellaValle97}  and $850\pm50$
pc \citep{Duerbeck92} , respectively.
\subsubsection{LW Ser}
\citet{Gehrzetal80} analysed the  expansion evolution of LW~Ser,
and  they adopted  $E(B-V)\sim0.32$ mag  corresponding to  the visible
maximum of expansion. \citet{CohenR83} also adopted $E(B-V)=0.32$ mag.
The reddening towards LW~Ser was also estimated by \citet{Prabhu87} as
$E(B-V)=0.52$ mag  from intrinsic  colour in  15th day  after maximum.
However the  interstellar reddening  for this  nova is  not determined
using spectroscopic  methods, we adopted  an arithmetic  mean of
photometric  estimates as  $E(B-V)=0.39\pm0.1$ mag,  and obtained  the
distance of LW Ser as  $d_\text{c}=1.27\pm0.40$ kpc.  The calculated distance
depends on the  first points in the reddening  curve, but interstellar
extinction  models of  \textsc{galaxia}  and Drimmel  are  consistent with  our
relation.  Thus, we conclude that  the distance is 1.27~kpc.  However,
our result is much smaller than 2.34 \citep{CohenR83} and 5~kpc
\citep{Gehrzetal80}  obtained from  expansion  parallaxes.  For  these
parallaxes, interstellar reddening has  to be 0.66 and 0.72~mag,
respectively.
\subsubsection{V382 Vel}
V382  Vel   is  a   very  fast   nova  identified   as  a   neon  nova
\citep{Woodwarde99}. In this paper, the interstellar reddening towards
V382 Vel was adopted to be $E(B-V)=0.12\pm0.03$ mag corresponding 
to  a mean  value of  the  reddening estimates  in the  literature;
$E(B-V)=0.05$ mag from various lines ratios and $E(B-V)=0.09$ mag from
Na~\textsc{i} D equivalent width both given by \citet{DV02} , $E(B-V)=0.20$ mag
from  flux  ratio  and  spectra  of V382  Vel  in  \citet{Shoree03}  ,
$E(B-V)=0.15\pm0.05$  mag   \citep{Hachisu14,  HachisuK16}   from  the
colour-colour evolution , $E(B-V)=0.2$  and 0.12 mag \citep{Hachisu14}
using  the  hydrogen  column  density given  by  \citet{MukaiI01}  for
various coefficients  of reddening-column density relation.   Thus, we
calculated the  distance of V382  Vel to be $d_\text{c}=0.9\pm0.8$  kpc. Note
that,  our estimation is  based on  the  first  points of  the
reddening-distance relation towards V382 Vel,  and errors of the colour
excesses obtained  from the RC stars  in the relation is  large. These
make the lower limit of the distance to be uncertain, and the error of
the  distance to  be large.   However, our  result is  consistent with
$0.8\pm0.1$  kpc \citep{Tomove15}  calculated from  a velocity  of the
expanding shell of about $1800\pm100$ km/s.
\subsubsection{LV Vul}
\citet{Fernie69} used  colour excesses  of B  stars towards  LV~Vul to
determine the reddening as  $E(B-V)=0.6\pm0.2$~mag which is consistent
with $E(B-V)=0.6\pm0.05$  mag estimated by \citet{Hachisu14}  from the
colour-colour evolution. \citet{Tempesti72}  adopted  an intrinsic
colour  of  $(B-V)_{0,\text{max}}=0.35$  mag  \citep{Schmidt57}  to
obtain $E(B-V)=0.55$ mag from the colour at maximum. The average value
of these two colour excesses,  $E(B-V)=0.57\pm0.05$ mag, placed LV Vul
at a distance of $d_\text{c}=4.09\pm0.22$ kpc using UKIDSS photometry.  Based
on  spectroscopic  analysis  of  LV  Vul,  expansion  velocities  were
measured to be between 900 and  2700 km/s indicating  the distance
0.92   \citep{Slavine95}   and   2.5   kpc   \citep{Dellavalle98},
respectively.  Note that, \citet{Slavine95} concluded that the remnant
is not  sufficiently resolved  to measure  its expansion  velocity and
angular size reliably. The distance  estimated in this study is larger
than  those  measured  from  the expansion  parallaxes.   It  is
possible that  the spherical  shell assumption and  poorly resolved
nova shell could  reason for the distance to be  calculated lower than
expected from the reddening value.
\subsubsection{NQ Vul}
The  reddening  towards  NQ~Vul  was  measured  as  $E(B-V)  =
0.7\pm0.2$ mag by  \citet{Younger80} from the equivalent  width of the
interstellar   band   at   6614~{\AA},   $E(B-V)   =   0.8$   mag   by
\citet{Yamashita77} from the  comparison of colours of  F super giants
with this nova, $E(B-V) = 0.9\pm0.3$ mag by \citet{MartinM77} from the
interstellar    polarization,   and    $E(B-V)   =    1.2\pm0.2$   mag
\citet{Carney77} from  the interstellar CH and  CH+ equivalent widths.
However \citet{HachisuK16} re-analysed  the colour-colour evolution of
NQ Vul, and they obtained the same reddening as $E(B-V) = 1.00\pm0.05$
mag as given in their previous study \citep{Hachisu14}.
In addition to these  estimates, \citet{KlareW78} used three different
methods  to  calculate  reddening  and   distance  of  NQ  Vul.   They
calculated    $E(B-V)=\    \leq0.56,\    0.58,\    0.9$    mag    from
polarization--$A_V$  relation,  blackbody--angular  diameter  relation
based on  IR measurements  and from  interstellar polarization,
respectively.  The  proper distance  of NQ Vul  was calculated  as $d=
1.16\pm0.21$ \citep{DD00}  and $1.6\pm0.8$ using  expansion parallaxes
\citep{Slavine95}. For a mean of reddening estimates except lower ones
given by  \citet{KlareW78}, $E(B-V)=0.92\pm0.2$ mag, the  distance was
calculated as  $d_\text{c}=2.0\pm0.5$ kpc  in this  study, which  is slightly
larger than the distance inferred from the expansion parallax.
\subsubsection{QV Vul}
QV  Vul  was spectroscopically  studied  to  obtain the  reddening  as
$E(B-V)=0.4\pm0.05$ mag by \citet{Scotte94} and as $E(B-V)=0.4$ mag by
\citet{Andreae94} both from  recombination lines of H and  He, and the
reddening map  \citep{NK80}. The reddening  was also determined  to be
$E(B-V)=0.60\pm0.05$ mag from  nova--giant sequence \citep{Hachisu14},
and $E(B-V)=0.32$ mag comparing photometric and spectroscopic behaviour
of QV  Vul \citep{Gehrze92}.  We  adopted the  reddening of QV  Vul as
$E(B-V)=0.4\pm0.05$  mag estimated  spectroscopically, and  calculated
the distance as  $d_\text{c}=1.2\pm0.3$ kpc, which is smaller than the
2.7 kpc \citep{DD00}  value obtained using an average  velocity of the
O~\textsc{iii}$\lambda$5007 and  N~\textsc{ii} $\lambda$6584 lines.  For  the distance of
2.7 kpc  \citep{DD00}, our  reddening-distance relation  indicates the
reddening of QV Vul as 0.52 mag.
%
\subsection{Sources without expansion parallax measurements}
In this  section, reddening measurements of  the sources without
an expansion parallax measurement is presented.

\subsubsection{CI Aql}
\citet{Ijima12}    derived    the     interstellar    extinction    as
$E(B-V)=0.92\pm0.15$ mag from the  average of six estimations obtained
using diffuse interstellar  absorption bands (DIB) and Na~\textsc{i}  D1 and D2
lines.   Their result  is  in agreement  with $E(B-V)=0.85\pm0.3$  mag
given  by \citet{Kissetal01}  within the  uncertainties, where  DIB at
5849 {\AA}, 6613 {\AA}, and Ca~\textsc{ii}  3934 {\AA} were used to calculate a
mean reddening value.  \citet{BurlakE01} analysed the Balmer decrement
to obtain  the interstellar reddening to  be $E(B-V)=0.91\pm0.11$ mag.
Besides  spectral estimations,  \citet{LederleK03}  found a  reddening
towards  CI  Aql as  $E(B-V)=0.98\pm0.1$  mag  using the  mean  values
adopted   in  the   pre-outburst   light  curve   model  \citep[][
  $E(B-V)=0.85$    mag,    $E(B-V)=1.00$    mag   in    the    revised
  version]{HachisuK01,HachisuK02}  and  reddening  estimates  obtained
from  the   spectral  analysis   in  the  literature.    In  addition,
\citet{Schaefer10} calculated  the reddening as $E(B-V)=0.8$  mag from
intrinsic colours at $t_2$.  All of these estimations are in agreement
with each other except with  $E(B-V)=1.5\pm0.15$ mag obtained from the
ratio of OI  lines \citep{Lynchetal04}, and $E(B-V)=0.2$  mag from the
intrinsic    colours   at    max   \citep{Schaefer10}.     Note   that
\citet{Schaefer10} used  results of \citet{Kissetal01}.   However, the
reason of the disagreement between results obtained from spectroscopic
methods  is  not  known.   We  adopted  a  mean  value  of  the
spectroscopic estimates except the value given by \citet{Lynchetal04},
$E(B-V)=0.92\pm0.22$ mag,  and calculated  the distance  of CI  Aql as
$d_\text{c}=3.0\pm0.5$ kpc.  In addition, its distance was also calculated as
$d_\text{c}=4.24\pm0.3$ kpc using $E(B-V)=1.5\pm0.15$ mag.  These results are
between $d=2.6\pm1.3$ \citep{Lynchetal04}  and $5.0^{+5.0}_{-2.5}$ kpc
\citep{Schaefer10}.
\subsubsection{V1301 Aql}
\citet{VanGenderen77} derived the reddening towards V1301 Aql from the
colour-colour  evolution and  concluded that  the reddening  cannot be
larger than  0.8 mag.   This value  is consistent  with that  given by
\citet{Vyrba75},  who adopted  $E(B-V)=0.61$  mag based on  the
reddening  of objects  near the  nova.   Using  this estimate  we
  calculated the distance of V1301~Aql as $d_\text{c}=2.2\pm0.3$ kpc.  On the
  other  hand, \citet{Miroshnichenko88}  calculated the  reddening as
$E(B-V)=0.9\pm0.06$ mag  using their stabilization stage  method. This
estimate  is larger  than the highest  reddening  in our  map,
corresponding to a distance of $d_\text{c}>7$ kpc.
\subsubsection{V1493 Aql}
Measurements of  the reddening and  distance towards this nova  have a
high  scatter.   First  of  all,  \citet{Bonifacioe00}  assumed
$E(B-V)=0.33\pm0.1$  mag  from  the  intrinsic  colour  of  novae  two
magnitudes below  maximum, and  calculated the distance  to be  in the
range 17.6-40.4~kpc  using three  different approaches.   They pointed
out that the nova should be on the boundary or outside the Galaxy.  On
the other  hand, \citet{Arkhipovae02}  compared hydrogen  Balmer lines
with  the theoretical  ones  to  obtain  a  reddening  of
$E(B-V)\sim1.56\pm0.3$   mag,   and   calculated   the   distance   as
$d=4.3\pm0.3$ kpc.  \citet{Venturinie04} used observed OI 8446, 11287,
and   13164  {\AA}   line   ratio  to   estimate   the  reddening   as
$E(B-V)=0.57\pm0.14$ mag.  Based on Della Valle's MMRD relation,
they  calculated  the  distance  to  be  $d=25.82\pm1.81$  kpc.
\citet{Munarie08b} investigated  the reddening towards V1493  Aql from
the colour evolution and estimates in the literature, and they adopted
the reddening  as 0.57 mag  given in \citet{Venturinie04}.   They also
pointed that the odd colour evolution reported by \citet{Bonifacioe00}
is  highly suspicious.   In addition,  \citet{HachisuK16} investigated
the colour-colour  evolution and reddening-distance relation  of V1493
Aql  and adopted  the  reddening as  $E(B-V)=1.15$  mag. Although,  we
calculated  the  distance  of  V1493 Aql  as  $d_\text{c}\geq4.2$  kpc  using
$E(B-V)=1.15\pm0.2$  mag, we  adopted  $E(B-V)=0.57\pm0.14$ mag  which
depends on  spectroscopic methods, and  conclude that the  distance of
this objects is $d_\text{c}=1.72\pm0.66$ kpc.
\subsubsection{V1721 Aql}
From a  comparison with  other novae at  a similar  early evolutionary
state, \citet{Heltone08}  obtained the reddening towards  V1721 Aql as
$E(B-V)=3.0$ mag, which implies $d_\text{c}=7.5\pm2.0$ kpc .
\subsubsection{OY Ara}
\citet{Zhao97} determined the interstellar reddening towards OY Ara as
$E(B-V)=0.32\pm0.06$ mag  using continuum spectra of  this nova. Since
there  is no  other reddening  estimation, we  adopted this  reddening
estimate and calculated the distance of OY Ara as $d_\text{c}=1.3\pm0.7$ kpc.
\subsubsection{V1065 Cen}
V1065  Centauri  (Nova   Cen  2007)  had  a   relatively  bright  nova
outburst, which  reached a maximum  brightness of
$m_V=7.6$  mag.    \citet{Heltone10}  carefully  analysed the
spectroscopic evolution of V1065 Cen  and obtained the reddening using
several methods: $E(B-V)=1.05\pm0.5$ mag from the equivalent widths of
Na ~\textsc{i} DI  line, $E(B-V)=0.29\pm0.07$  and $0.45\pm0.06$  mag from  the
colour index  at maximum  and at  $t_2$, and  $E(B-V)=0.79\pm0.01$ mag
from the flux ratio of $H_\alpha/H_\beta$ for the first 160 days
after outburst.  Although  the equivalent widths of Na~\textsc{i}  lines give a
precise measurement of reddening,  \citet{Heltone10} did not use these
equivalent widths,  since the  spectral resolution  was not  enough to
decide  whether these  features  are saturated  or  not.  Hence,  they
adopted a mean value  of reddening as $E(B-V)=0.5\pm0.1$ mag and
calculated the  distance to be  $8.7$ kpc  from the MMRD  relations of
\citet{DellaValle97}.  They  also noted that the  derived distance for
V1065   Cen    may   not   be   well    constrained.    In   addition,
\citet{HachisuK16} inferred the reddening towards  V1065 Cen as
$E(B-V)=0.45\pm0.05$ mag using the intrinsic colour evolution of
the nova,  which is consistent  with that given  by \citet{Heltone10}.
Therefore, we used the mean value of the two reddening estimates
given     by    \citet{Heltone10}     and    \citet{HachisuK16}     as
$E(B-V)=0.47\pm0.05$  mag, and  calculated the  distance to  be
$d_\text{c}=3.3\pm0.5$ kpc.  Note that since reddening curve does not clearly
exceed  the limits  of the  given reddening,  the upper  limit of  the
distance is slightly uncertain.
\subsubsection{V868 Cen}
The  interstellar  reddening towards  V868  Cen  was estimated  to  be
$E(B-V)=1.72\pm0.1$ mag  as an  average value  of the  following
results; 1.7  mag \citep{Orio01}  from the  column density  of neutral
hydrogen and 1.63$-$1.87~mag \citep{Williams94} from the H~\textsc{i} and He \textsc{ii}
recombination lines.  Using this  average reddening, we calculated the
distance of V868 Cen as $d_\text{c}=5.2\pm0.4$ kpc.
\subsubsection{AR Cir}
\cite{Tapea13}   investigated  spectra   of   AR   Cir,  and   assumed
$E(B-V)\sim2$ mag after testing different values.  This estimate
is not in agreement with \cite{BE94}, who adopted $E(B-V)=0.78$ mag by
comparing reddening measurements from \cite{NK80}  reddening map
  corresponding   to    magnitudes   and   distances   of    AR   Cir
\citep{DuerG93}.   Although   a  more  reliable  measurement   of  the
reddening from the analysis of interstellar lines is required, as also
mentioned  by  \cite{Tapea13},  we  used  $E(B-V)=2.0\pm0.4$  mag  and
calculated the distance of this nova as $d_\text{c}=3.15\pm0.56$ kpc.
\subsubsection{BY Cir}
\citet{Greeleye95}  concluded  that  the  spectral  lines  at  shorter
wavelengths  show significant  extinction,  and  they corrected  their
fluxes using a reddening of $E(B-V)=0.11$ mag based on the He \textsc{ii}
1640/ He \textsc{ii}  1085 line ratio.  However,  \citet{Greeleye95} noted that
this reddening is only an  approximation, since their corrected fluxes
strongly depend  on the  actual extinction curve.   Additionally, they
used the intrinsic colour at outburst maximum and obtained
 a reddening of $E(B-V) = 0.15\pm0.06$ mag. They further analysed
their He \textsc{ii} flux with the presence of N \textsc{ii} airglow emission which
  resulted   in   an   upper   limit   of    $E(B-V)\leq0.56$   mag.
\citet{Evanse02} used polarization in line  of sight of this nova, and
obtained  a  distance  for  BY  Cir  greater  than  2$-$2.5  kpc  with
$E(B-V)\geq0.55$.  Hence, we conclude  that the reddening towards
BY  Cir  is  uncertain,  since   the  discrepancy  between the
estimation  of   \citet{Greeleye95}  and  \citet{Evanse02}   is  high.
Considering all  the estimations mentioned above,  we calculated three
distances  as  $\leq0.95$,  $1.0\pm0.2$ and  $\geq3.25  (-0.68)$  kpc.
Since our reddening-distance relation is  in very good agreement
with Drimmel and \textsc{galaxia} extinction  curve, we adopted the distance as
$d_\text{c}=1.0\pm0.2$  kpc  from the  first  points  in our  relation  using
$E(B-V)=0.13\pm0.06$ mag, which is the mean value of results in
\citet{Greeleye95}.
\subsubsection{V394 CrA}
\citet{Schaefer10} used a  reddening value  towards V394  CrA as
  $E(B-V)=0.2\pm0.2$  mag  by   comparing  their  reddening  estimates
  obtained from  the relation between observed  colours and intrinsic
colours at  the outburst  maximum and at  $t_2$ with  $E(B-V)=0.2$ mag
given by  \citet{Duerbeck88} from  the reddening  of objects  near the
nova.   Since the  reddening given  by \citet{Schaefer10}  has a  very
large error, we used the  reddening of $E(B-V)=0.2$ mag and calculated
an uncertain distance as $d_\text{c}=2.0$ kpc.
\subsubsection{Q Cyg}
The   interstellar   reddening   towards   Q~Cyg   was   obtained   as
$E(B-V)=0.26\pm0.06$    mag    from    the    2200    {\AA}    feature
\citep{SelvelliG13},  which is  lower than  previous estimates  in the
literature \citep[see][]{SelvelliG13}.  This estimate corresponds to a
distance of $d_\text{c}=1.22\pm0.9$ kpc.
\subsubsection{V2274 Cyg}
\citet{Rudyetal03} estimated  the reddening to  be $E(B-V )  =1.30 \pm
0.2$ mag from the fluxes of three  O~\textsc{i} lines, which is consistent with
$E(B -V  ) =  1.35 \pm  0.1$ mag  \citep{Hachisu14} obtained  from the
colour-colour evolution of V2274 Cyg.  They also used Paschen lines to
obtain $E(B-V)=1.47\pm0.15$ mag.  We adopted the mean of the reddening
estimates except  the  one obtained  using Paschen  lines, since
the reddening does not  climb  up to  1.47  mag in  our
reddening-distance relation  constructed  using  the RC  stars,
which extends up to 13  kpc (based on 2MASS  photometry. 
  The  adopted   reddening value  indicates  a  distance  of
$d_\text{c}=9.0\pm2.0$ kpc  for V2274~Cyg,  which is  obtained from  the last
points in the reddening curve.
\subsubsection{V2275 Cyg}
\citet{Kissetal02}  express that  the  interstellar reddening  towards
V2275~Cyg is well-determined using various  methods and its mean value
was given as  $E(B-V)=1.0\pm0.1$ mag.  Thus, we adopted  this value as
the colour excess of V2275  Cyg, which indicates $d_\text{c}=3.62\pm0.47$ kpc
using our distance estimation method.
\subsubsection{V2467 Cyg}
\citet{Steeghse07} estimated  the interstellar reddening  towards V2467
Cyg as  $E(B-V)=1.0-1.5$ mag  using the brightness  and colour  of the
progenitor,  while \citet{Munarie07c}  used  equivalent  width of  the
Na~\textsc{i}~D2  to obtain  the  reddening as  $E(B-V)=0.31$  mag.  Using  O~\textsc{i}
lines, the reddening of V2467~Cyg was  estimated as 1.5 and 1.7 mag by
\citet{Mazuk07}  and  \citet{Russelle07}. Note  that,  \citet{Mazuk07}
concluded that  the  evidence of dust  formation does not
exist in their  spectra.  In addition, the reddening  was estimated as
$E(B-V)=1.16\pm0.12$ mag \citep{Poggiani09}  using intrinsic colour at
$t_2$,  and   $E(B-V)=1.38\pm0.12$  mag  \citep{Shugarove10}   from  a
comparison between the observed and absolute magnitudes in maximum and
interstellar  extinction.   Finally, \citet{HachisuK16}  obtained  the
reddening as  $E(B-V)=1.40\pm0.05$ mag comparing intrinsic  colours of
V2467 Cyg with  similar novae.  All of these  estimates are consistent
with each  other except the one given by  \citet{Munarie07c}. We
  also  did not  take  this reddening  measurement  into account  when
  calculating the  mean reddening as $E(B-V)=1.40\pm0.2$  mag.  Using
this value, we calculated the distance of V2467 Cyg as $d_\text{c}=1.5\pm0.3$
kpc.
\subsubsection{V2468 Cyg}
Using  O~\textsc{i}  lines,  \citet{Rajetal15} estimated  the reddening  toward
V2468~Cyg to  be $E(B-V)=0.77\pm0.15$ mag  which is in  agreement with
\citet{Rudye08}.  Similar  values were also calculated  from different
methods; 0.8  mag \citep{Schwarzetal11b,  Tarasova13} from  the Balmer
decrement method, $0.8\pm0.1$  mag \citep{IijimaN11}  using the
relation  between the  interstellar  reddening and  column density  of
hydrogen  atoms converted  from Na,  0.78 mag  \citep{Chochole12} from
intrinsic colour at $t_2$, and $0.75\pm0.05$ mag \citep{HachisuK16} by
comparing  intrinsic colour  evolution of  V2468 Cyg.   Thus, we  used
$E(B-V)=0.78\pm0.1$  mag  as the  mean  reddening,  which indicates  a
distance of $d_\text{c}=6.8\pm1.0$ kpc.
\subsubsection{V2491 Cyg}
The classical  nova V2491~Cygni  was classified  as an  extremely fast
nova \citep{Tomove08a} based on  the observed rate of decline of
$t_2\sim4.6$ days  \citep{Tomove08b}.  The interstellar  reddening has
been  estimated  from   the  O ~\textsc{i}  line  ratio   as  $E(B-V)=0.3$  mag
\citep{Lynche08}.  \citet{Rudyet08} used the  same lines and concluded
that  the  reddening  to  be  $E(B-V)=0.43$  mag.   \citet{Munarie11b}
studied the  photometric and  spectrometric evolution of  V2491~Cyg in
detail,  and adopted  $E(B-V)=0.23\pm0.01$  mag based on  three
measurements: $E(B-V)=0.24$~mag  from the equivalent  widths of
the Na~\textsc{i} line, and $E(B-V)=0.23$ and $E(B-V) =0.22$ mag from intrinsic
colour  of  novae at  maximum  and  $t_2$, respectively.   We  adopted
$E(B-V)=0.23\pm0.01$ mag \citep{Munarie11b}, which was also adopted by
\citet{HachisuK16} comparing colour evolution curves of V2491 Cyg, and
calculated the distance as $d_\text{c}=2.1\pm1.4$ kpc.
\subsubsection{V407 Cyg}
V407 Cyg had been classified as a symbiotic star, until when the first
recorded explosive mass  ejection in 2010 resembled  a classical novae
\citep{Maehara10}.  It  is now  known as  a symbiotic  recurrent novae
with  a  RG secondary  \citep{Munea2011}.   \citet{Munarie90}
obtained  the  reddening  as  $E(B-V)=0.57$  mag  based  on  the
  modelling    of   the  spectral   energy    distribution,   while
\citet{Shoree11} found  $E(B-V)=0.45\pm0.09$ mag from the  analyses of
DIBs, but adopted  $E(B-V)=0.5\pm0.05$ mag based  on a comparison
  of different  reddening values  for different spectral  types, DIB,
and SED.   These results are consistent  with $E(B-V)=0.61\pm0.05$~mag
\citep{Iijima12} obtained from the intrinsic colour of V407 Cyg.
Thus, we used $E(B-V)=0.55\pm0.05$ mag for V407 Cyg and calculated its
distance as $d_\text{c}=3.50\pm0.30$ kpc.
\subsubsection{IM Nor}
IM   Nor  is   a  recurrent  nova,   and  was   investigated  by
  \citet{Schaefer10} in detail.  Although, \citet{Schaefer10} gave two
  reddening  estimates as  0.5 and  0.7 mag  from intrinsic  colour at
  maximum and $t_2$, respectively,  they adopted the reddening towards
  IM Nor as $0.8\pm0.2$ mag,  which is consistent with $E(B-V)\geq0.8$
  mag \citep{Duerbecke02} and  the range of 0.5-1.1  mag determined by
  \citep{Orio05}.  We calculated the distance of IM Nor as $2.4\pm0.8$
  kpc    using   the    crude    reddening    estimate   adopted    by
  \citet{Schaefer10}.
\subsubsection{V382 Nor}
The interstellar  reddening towards  V382 Nor is  not well-determined.
\citet{Nesse07}  used  a Swift  X-ray  observation,  and obtained  the
reddening to be  $E(B-V)=0.6-1.1$ mag from  the inferred hydrogen
  column density.  This value is consistent with  that found by
  \citet{Ederoclitee05} whose  conclusion shows  that  the saturated
Na~\textsc{i} interstellar lines refer to a high reddening.  Since there
is no  other estimation for the  reddening of V382 Nor,  we calculated
its  distance to  be within  3$-$5 kpc  using  the  given
reddening  range,  and  $4.0\pm0.53$  kpc   from  the  mean  value  of
$0.85\pm0.15$ mag.
\subsubsection{V2214 Oph}
\citet{Lynchetal89} adopted  the interstellar reddening  towards V2214
Oph  to   be  $E(B-V)=0.73\pm0.13$  mag,  which   is  consistent  with
$E(B-V)=0.74\pm0.07$   mag  \citep{Erwinetal92}   obtained  from   the
hydrogen line ratios. In addition,  \citet{Williams94} used H~\textsc{i} and He
\textsc{ii} recombination lines  to estimate similar result  as $\sim0.65$ mag.
Using  $E(B-V)=0.73\pm0.13$  mag,  we  determined  the  most  probable
distance as $d_\text{c}=2.6\pm1.0$ kpc.
\subsubsection{V2295 Oph}
\citet{DV94}  investigated the  interstellar absorption  towards V2295
Oph.  First, they used the equivalent width of Na~D~lines, which
imply  a   reddening  of   $E(B-V)=0.75\pm0.2$   mag,   while
$E(B-V)=0.6-0.9$ mag was estimated from interstellar absorption bands.
They adopted $(B-V)_{0,\ max} =-0.1$ mag  to obtain a colour excess of
$E(B-V)=1.0$ mag,  which is consistent with  $E(B-V)=1.0$ mag obtained
from  the  hydrogen column  density  \citep{DV94}.   Using the  column
density  of  the atomic  hydrogen  at  21 cm,  \citet{DV94}  estimated
$E(B-V)=1.0$  mag.  Finally,  they determined  $E(B-V)=1.15$ mag  from
the Balmer decrement. Based on the estimations summarized
above, \citet{DV94} adopted the reddening as $E(B-V)=0.9\pm0.2$ mag
towards V2295 Oph.  Using  $E(B-V)=0.9\pm0.2$ mag, the distance should
be greater than 1.4 kpc, but the most possible distance was
  determined as 2.1  kpc.  We were able to calculate  the distance to
be $1.64\pm1.30$  kpc  using $E(B-V)=0.75\pm0.2$  mag.  Note that
in this line of sight the reddening is nearly constant at $E(B-V)=1.0$
mag for distances larger than 2.5 kpc.
\subsubsection{V2487 Oph}
From the observations  of the O~\textsc{i} at  $\lambda8446$ and $\lambda11287$
lines,  \citet{Lynchetal00}   estimated  the   colour  excess   to  be
$E(B-V)=0.38\pm0.08$   mag,    indicating  a    distance   of
$d_\text{c}=1.08\pm0.28$  kpc,  which was  also  adopted  in our  study.   In
addition, \citet{Schaefer10}  used intrinsic colours at  outburst
maximum and $t_2$  to calculate the reddening towards  V2487 Oph as
$E(B-V)=0.4$  and $0.6$  mag.   These  reddening values  indicate
  distances  of $1.12\pm0.33$  and  $1.91\pm0.90$ kpc,  respectively.
Interestingly, the distance for this object as estimated from the
  MMRD  relation  is  found  to  be in  the  range  12-27.5  kpc
\citep{Schaefer10, Lynchetal00}.   However, our results show  that the
distance of V2487 Oph should be 1-2 kpc.
\subsubsection{V2576 Oph}
The  interstellar   reddening  towards  V2576  Oph   was  reported  by
\citet{Russelle06}  as  0.62 mag  from  infrared  spectra. Using  this
value, we calculated $d_\text{c}=1.8\pm1.0$ kpc.
\subsubsection{V2615 Oph}
The  reddening   prior  to  dust   formation  was  determined   to  be
$E(B-V)=$1.0-1.3   mag    using   the   strength   of    O  ~\textsc{i}   lines
\citep{Russell07b}.   \citet{Munarie08}  calculated $E(B-V)=0.89$  and
0.91  mag based on  assumed  intrinsic colours  at maximum  and
$t_2$, and  showed that the  equivalent widths of  interstellar Na~\textsc{i}~D
lines imply  a lower limit  of reddening  to be $E(B-V)>0.7$  mag.  In
addition, \citet{HachisuK16}  compared intrinsic  colour of  V2615 Oph
with those of FH Ser, QV Vul, and V705 Cas, and obtained the reddening
as  $E(B-V)=0.95\pm0.05$  mag.   We  adopted  a  median  reddening  as
$E(B-V)=0.95\pm0.15$ mag  and calculate the  distance of V2615  Oph as
$d_\text{c}=2.1\pm0.8$ kpc.
\subsubsection{V2672 Oph}
V2672 Oph, which  is an extremely fast nova, was  discovered at almost
the maximum brightness.  \citet{Munarie11} adopted an average value of
the  photometric  and  spectroscopic  estimates of  the  reddening  as
$E(B-V)=1.6\pm0.1$  mag. Using  $E(B-V)=1.6\pm0.1$~mag, we  calculated
the distance of this object as $d_\text{c}=3.12\pm0.69$~kpc.  Note that
the distance strongly  depends on the assumed  reddening for the
direction of this nova, since the reddening evolution is shallow
  for distances larger than $\sim2.5$ kpc.
\subsubsection{V2674 Oph}
\citet{Munarie10b} obtained  $BVR_\text{c}I_\text{c}$ photometry  of V2674  Oph, and
determined  $V_{max}=9.4$   mag,  and  $t_2=18$  and   $t_3=31$  days.
\citet{Munarie10b} also  calculated  the reddening  towards V2674
Oph  to be  $E(B-V)=0.7\pm0.1$  mag using  the  mean intrinsic  colour
$(B-V)_0=+0.23\pm0.06$   mag    at   the   time   of    maximum,   and
$(B-V)_0=-0.02\pm0.04$ mag at  $t_2$ \citep{vandenbergY87}. Using this
reddening,  the  distance  of this  source was  calculated  as
$d_\text{c}=1.65\pm0.38$ kpc.
\subsubsection{V972 Oph}
In order  to use  in the  distance estimation,  we adopted  the colour
excess of V972 Oph as $E(B-V)=0.5\pm0.1$ mag \citep{ZwitterM96}, which
was derived  from the  strength of the  Na~\textsc{i}~D absorption  lines. Note
that there is no other reddening  estimate reported in the
literature.   We   calculated  the   distance  of   V972  Oph   to  be
$d_\text{c}=0.87\pm0.15$ kpc.  Since this distance estimation  depends on the
first points in the reddening-distance relation towards V972 Oph,
the lower limit of the distance remains uncertain.
\subsubsection{V529 Ori}
Two interstellar reddening estimates  have been reported towards
the     nova     V529      Ori.      \citet{RingwaldN97}     obtained
$E(B-V)=0.32\pm0.02$~mag from the visual component  of V529~Ori
under  the assumption  that it is  a  K7 $-$  M0 dwarf,  while
\citet{Schmidtobreicke05} used  the optical  continuum and  flux ratio
H$_\alpha$/H$_\beta$  to determine  $E(B-V)=1.5$  mag.  These  results
indicate  very different  distances, $d_\text{c}=1.29\pm0.35$  and $d_\text{c}\geq7$
kpc, respectively. Note that based  on the RC stars the reddening
  increases up to  $E(B-V)\sim0.8$ mag at the distance  of $\sim7$ kpc
  in    our    study.     The    reddening    estimate    given    by
\citet{Schmidtobreicke05} seems to be  over-calculated or the distance
of V529  Ori is  really  at greater  distances, while  the other
method  used  to  determine  the   reddening  is  not  very  reliable.
\citet{Schmidtobreicke05}  concluded that  several  neutral or  easily
excitable elements  should be seen  in the  their spectra due  to high
reddening.  However,  since no such  lines are present,  they proposed
that  V529  Ori   is  a  T  Tauri  star.   Thus,   the  reddening  and
identification  of V529  Ori  as a  nova is  doubtful, and  this
object should be investigated in detail.
\subsubsection{V1187 Sco}
Using  the  OI  lines,  \citet{Lynchetal06}  derived  a  reddening  of
$E(B-V)=1.56\pm0.08$  mag, which  is  consistent  with their  previous
estimation of $E(B-V)=1.2-1.5$ mag extracted from equivalent widths of
the Na~\textsc{i}~D lines.  \citet{Lynchetal06}  also analysed old photographic
plates to obtain  the colour of V1187 Sco in outburst.  Intrinsic
colour  method indicates  a reddening  of $E(B-V)=1.3$  mag, which  is
consistent  with their  previous  determinations.  Thus,  we used  the
colour excess of this nova  as $E(B-V)=1.56\pm0.08$ mag and calculated
its distance to be $d_\text{c}=3.20\pm0.12$ kpc.
\subsubsection{V1309 Sco}
\citet{Tylendaetal11} recently showed that V1309 Sco is the first merger between two convective stars to ever be observed
and Nova Sco 2008 have resulted from the merger of the two stars in the contact binary. However, V1309 Sco is untypical nova, 
we analysed its reddening estimates and calculated its distance.
 \citet{Masonetal10} calculated the  average equivalent width of
 Na K~\textsc{i} $\lambda7699$ interstellar line from 18 different spectra, and
 determined the  reddening to  be $E(B-V)=0.55$ mag  for V1309  Sco in
 outburst. \citet{Tylendaetal11} adopted $E(B-V)=0.8$ mag by comparing
 $BVRI$  photometry   and  spectral   type  of   the  secondary
 star. Because it is based on  a more reliable method, we here adopted
 $E(B-V)=0.55\pm0.05$  mag  \citep{Masonetal10}   and  calculated  the
 distance of this object as $d_\text{c}=2.5\pm0.4$ kpc.
\subsubsection{V1324 Sco}
Nova  Sco 2012  (V1324  Sco)  was discovered  in  May  2012 while  its
outburst    had   just    began.    Thus,    its   light    curve   is
well-determined. \citet{Munarie15} used intrinsic  colour of V1324 Sco
at  the  outburst  max  and  $t_2$,  and  obtained  the  reddening  as
$E(B-V)=1.21$ and  $1.25$ mag, respectively.  They  adopted an average
of 1.23  mag, which  is good  agreement with  $E(B-V)=1.23\pm0.15$ mag
derived by \citet{Finzelle15} from total equivalent widths of the Na \textsc{i}
D  and  KI absorption  lines.   \citet{Finzelle15}  also used  another
independent  method   of  equivalent  widths  of   DIBs  to  determine
$E(B-V)=1.07\pm0.20$ mag, and adopted $E(B-V)=1.16\pm0.12$ mag.  Since
reddening  estimates  given  by   \citet{Finzelle15}  depend  on  high
resolution spectra of V1324  Sco, we adopted $E(B-V)=1.16\pm0.12$ mag,
which  results in  a distance  of $d_\text{c}=4.3\pm0.9$  kpc. Note  that our
result depends on the last point of reddening curve, and any reddening
estimate higher than  $\sim$1.4 mag will change  the distance of
V1324 to a lower limit of $d_\text{c}\geq 5$ kpc.
\subsubsection{V745 Sco}
\citet{Orioe15} investigated the X-ray spectrum of the recurrent
nova  V745 Sco,  and inferred  the hydrogen  column density  as
$N_H=(6.9\pm0.9)  \times  10^{21}$  cm$^-2$, which  corresponds  to  a
reddening  of $E(B-V)=1.0$  mag using the  relationship between
reddening  and hydrogen  column density  as $E(B-V)=N_H  / 6.8  \times
10^{21}$   cm$^-2$   \citep{Guvere09}.   However,   \citet{HachisuK16}
obtained  the reddening  as $E(B-V)=0.83\pm0.1$  mag using  the column
density   and  reddening   relation  given   by  \citet{Orioe15}   and
\citet{Liszt14},  respectively. But,  they  adopted $E(B-V)=0.70$  mag
from the  comparison of  their colour-colour diagram  of V745  Sco and
reddening maps.   We adopted a  mean value of the  reddening estimates
given above  as $E(B-V)=0.84\pm0.15$ mag, and  calculated the distance
of  V745  Sco  to  be  $d_\text{c}=3.5\pm0.85$ kpc.   In  addition  to  these
estimates, \citet{Williams94}  used H~\textsc{i} and He~\textsc{ii}  recombination lines
and  obtained   higher  reddening  values   as  1.22  and   1.13  mag,
respectively, which results in distances larger than 4.5 kpc.
\subsubsection{V977 Sco}
\citet{Williams94}  adopted  $E(B-V)=1.05\pm0.09$~mag obtained  as  an
average value of reddening estimates in range of 0.97-1.31 mag, 
which  were  derived  from  HI  and  H~\textsc{ii}  recombination  lines,  while
\citet{Andreae94} used  the  same spectral lines  and calculated
the reddening to be $E(B-V)=0.8$ mag.   We used the mean of these
  results as $E(B-V)=0.92\pm0.10$ mag, and calculated the distance to
be $d_\text{c}=3.3\pm0.6$ kpc.
\subsubsection{V992 Sco}
Interstellar  reddening  towards  V992  Sco  was  calculated  only  by
\citet{Williams94}  as $E(B-V)=1.33$  and 1.26  mag from  the H~\textsc{i}  and
He~\textsc{ii} lines,  respectively.  We used $E(B-V)=1.3\pm0.1$  mag to obtain
$d_\text{c}=2.63\pm0.35$ kpc.
\subsubsection{V368 Sct}
The Nova Scuti 1970 (V368 Sct) was discovered at  a brightness of
V  = 6.9  mag.  Although  V368 Sct  is an  old nova,  the interstellar
reddening in the direction of this  system has been studied only once.
\citet{Ciati74}    investigated   V368    Sct   photometrically    and
spectroscopically, and  estimated the  reddening as  $E(B-V)=0.5$ mag.
However  they  stated that  this  colour  excess is  uncertain.
  Therefore, the  distance calculated as $d_\text{c}=1.94\pm0.95$  kpc in our
  study is also uncertain.
\subsubsection{V443 Sct}
The  interstellar reddening  towards  V443 Sct  has  been examined  in
detail.   \citet{Anupamae92}  adopted $E(B-V)  =  0.4$  mag from  four
estimations: $E(B-V)  = 0.41\pm0.03$ mag from  the Balmer/Paschen line
ratios, $\geq0.07$  mag from polarization, $\geq0.25$  mag from colour
index  of  the expansion  light  curve,  and  $\leq0.9$ mag  from  the
comparison  with   globular  cluster  NGC   6705.   \citet{Williams94}
obtained the reddening of V443 Sct to  be 0.57$-$1.1 mag from H and He
recombination lines, which indicate a  mean reddening as 0.85 mag, but
\citet{Andreae94} obtained $E(B-V)=0.30$  mag from recombination lines
of  H   and  He.   Finally,  \citet{Hachisu14}   estimated  $E(B-V)  =
0.40\pm0.05$ mag  from the colour-colour  evolution of V443  Sct, with
respect  to  only several  data  points  after optical  maximum.   The
reddening    estimate   given    by   \citet{Williams94}    could   be
over-estimated, since  it is much  higher than the  maximum value
  reached in  our reddening curve for  the direction of V443  Sct and
other  reddening  estimates  given  above.  Thus,  we  used  a  median
reddening  as $E(B-V)  =  0.40\pm0.05$ mag  without taking  into
  account  the one  given by  \citet{Williams94}, and  calculated the
distance of V443 Sct as $d_\text{c}=1.4\pm0.5$ kpc.
\subsubsection{V476 Sct}
We adopted the interstellar reddening as $E(B-V)=1.9\pm0.1$ mag, which
is  the mean  value of  the reddening  estimates of  1.8 mag  from the
intrinsic  colour  method  at  $t_2$ \citep{Munariel06}  and  2.0  mag
indicated    by   O~\textsc{i}   lines    \citep{Perrye05}.    Note    that,
\citet{Munariel06} concluded that  the equivalent width of  the DIB at
6614 {\AA}  implies a massive  reddening affecting the nova.   For the
direction  of V476  Sct, we  obtained  a heavily  reddened UKIDSS  CMD
indicating a distance of $d_\text{c}=11.3\pm1.9$ kpc, which locate the system
beyond the edge of bulge.
\subsubsection{V496 Sct}
\citet{Raje12} obtained the interstellar reddening towards V496 Sct as
$E(B-V)=0.57\pm0.06$ mag from intrinsic colour of the nova at outburst
maximum of  outburst, and  $E(B-V)=0.65$ mag  using interstellar  Na \textsc{i}
line.   The reddening  was  also determined  as  $E(B-V)=0.50$ mag  by
\citet{HachisuK16} from  intrinsic colour  evolution of V496  Sct. The
reddening estimates are  in  the range of  0.50$-$0.65 mag, which
indicate  that the distance  is between 2.5 and 2.9 kpc 
  as calculated from the reddening-distance relation towards V496 Sct
in  this  study.   For  a  mean  value  of  the   estimates  given  by
\citet{Raje12}  and  \citet{HachisuK16}, $E(B-V)=0.57\pm0.7$  mag,  we
calculated the distance of V496 Sct as $d_\text{c}=3.2\pm0.8$ kpc.
\subsubsection{MU Ser}
The interstellar reddening towards MU Ser is not well-known.  There is
only  one estimate:  $E(B-V)=0.4\pm0.1$  mag which  was determined  by
\citet{Wargau83} from the UV continuum behaviour. Using this reddening,
we  calculated the  distance as  $d_\text{c}=0.72\pm0.53$ kpc.  However, this
calculation  depends on  the  first point  of reddening-distance
relation  although   it  is  consistent  with   the  theoretical  ones
\citep{Drimext,Sharma11}.
\subsubsection{WY Sge}
\citet{Kenyon88} investigated infrared colours,  spectral type of 
  the secondary, and visual absolute magnitude at max, and found
 the  reddening to  be in range of  0.32$-$1 mag, which  is the
same as  found by  \citet{Somers96}, who  calculated the  reddening as
0.39-1  mag  by  comparing  observed  colours  of  WY  Sge  with
  theoretical ones. A simple  average of these estimates, 0.67 mag,
is consistent  with the  generally adopted reddening of  0.6 mag
in other studies \citep{HarrisonG88,Orio01,Shafter97}.  However,
this mean  value is not in  agreement with the  $E(B-V)=1.6$ mag
\citep{SharaM03}  obtained from  the  strong  4428 {\AA}  interstellar
absorption  feature.   We  calculated  the   distance  of  WY  Sge  as
$3.1\pm0.3$  kpc  and $4.2\pm0.4$  kpc  using  reddening estimates  as
$0.67\pm0.13$ and $1.6\pm0.3$ mag, respectively.
\subsubsection{V1172 Sgr}
\citet{ringwaldetal96} investigated an optical spectrum of V1172
Sgr, and  calculated  the reddening as $E(B-V)=2.1\pm0.5$
mag. \citet{Weighte94}  argued that  V1172 Sgr  should be  a recurrent
nova, although this  reddening locate the optical colour  of V1172 Sgr
in the region where the systems  with dwarf secondaries on the CMD are
found.  Thus, they assumed a reddening value that satisfies
  the recurrent nova expectation as $E(B-V)=0.4$ mag.  Interestingly,
\citet{ringwaldetal96}    even   used    the   reddening    given   by
\citet{Weighte94}    for    their    analysis.     Furthermore,    our
reddening-distance relation obtained from  2MASS CMD towards V1172 Sgr
indicates that  reddening may not  be larger  than 1.4 mag.   Thus, we
calculated the distance of V1172 Sgr to be $d_\text{c}=0.86\pm0.14$ kpc using
$E(B-V)=0.4\pm0.08$ mag.
\subsubsection{V4077 Sgr}
\citet{Drechsele84}  de-reddened IUE  UV  spectra of  V4077 Sgr  using
$E(B-V)=0.3\pm0.1$      mag,      while     \citet{Mazehe85}      used
$E(B-V)=0.35\pm0.1$ mag  based  on a  measurement from  the 2200
      {\AA} feature.  These colour excesses  are in agreement with the
      0.32  mag given  by  \citet{Warner95}.  We  used the  mean
        reddening  of  V4077  Sgr  as  $E(B-V)=0.32\pm0.03$  mag  and
      calculated the distance of this source as $d_\text{c}=1.6\pm1.0$ kpc.
\subsubsection{V4160 Sgr}
\citet{Schwarzel07}  corrected   the  spectra   of  V4160   Sgr  using
$E(B-V)=0.6$ mag obtained from  the expansion light curve similarities
with V838 Her.   \citep{Masonel02} estimated $E(B-V)=0.35\pm0.04$
  mag  using  the  Na ~\textsc{i}  equivalent  width  measurement  due  to  the
  contamination  of nearby  He~\textsc{i}  emission.  However,  our reddening-
distance relation shows that the reddening cannot reach up to 0.6
  mag towards V4160  Sgr for distances smaller than 5  kpc.  Thus, we
conclude   that   the   distance    is   $d_\text{c}=1.8\pm1.2$   kpc   using
$E(B-V)=0.35\pm0.04$ mag.
\subsubsection{V4169 Sgr}
\citet{Scotte95} obtained  $E(B-V)=0.35\pm0.05$ mag and 0.33  mag from
      the  2200  {\AA}  feature and  optical  nebular  spectra,
      respectively,  while  \citet{Williams94}  used  H~\textsc{i}  and  He  \textsc{ii}
      recombination lines to  estimate the reddening as  0.41 and 0.51
      mag,   respectively.    Additionally,   \citet{Rossano94}
      calculated the  reddening to be $E(B-V)=0.35$  mag towards V1469
      Sgr from two  different spectral lines: $E(B-V)=0.4\pm0.2$
      mag from O~\textsc{i} lines and  $0.29\pm0.1$ mag from the Paschen lines.
      They concluded that the estimated reddening changes depending on
      the observation date and blend  effect. Since the spectral lines
      can be affected  by blending and dust formation  around the nova
      as     mentioned     in    \citet{Rossano94},     we     adopted
      $E(B-V)=0.35\pm0.05$  mag,  which  is consistent  with  previous
      results.  Based  on this value we  calculated the distance
      of the nova as $d_\text{c}=1.9\pm0.7$ kpc.
\subsubsection{V4332 Sgr}
Comparison of the spectroscopic  and photometric observations of V4332
Sgr   with    giants   template,    mainly   M   giants,    have   led
\citet{Martinietal99},   \citet{Tylendaetal05},  \citet{Kimeswenger06}
and \citet{Tylendae15}  to conclude  that the reddening  towards V4332
Sgr is $E(B-V) = 0.32 \pm 0.02$, $0.32 \pm 0.10$, $0.37 \pm 0.07$, and
$0.35$ mag, respectively.  In addition, \citet{Kaminaskie10} used
  a more  reliable reddening  estimation method based on the
equivalent width of Na~\textsc{i}~D lines and determined the reddening as
$E(B-V)=0.45\pm0.05$  mag, although  the  Na~\textsc{i}~D  lines were  probably
saturated.   They  used   column  density  of  H~\textsc{i}   by  neglecting  a
contribution  from molecular  hydrogen to  obtain $E(B-V)  = 0.22  \pm
0.04$ mag, and adopted $E(B-V) =0.32$ mag.  Using $E(B-V)=0.32\pm0.07$
mag  as a  reasonable  value between  the  above estimates,  we
calculated  the  distance as  $d_\text{c}=1.14\pm1.0$  kpc.   Note that,  the
estimates  from column  density of  H~\textsc{i}  line indicate  a distance  of
$d_\text{c}=0.75$ kpc.
\subsubsection{V4444 Sgr}
The difference  in the reddening  estimates reported for  this object,
cause a  scatter in  our distance  calculations.  \citet{Venturinie02}
determined  the  reddening as  $E(B-V)=1.07  \pm  0.10$ mag  from  the
observed  O~\textsc{i}  flux ratios.   Using this  estimate, we  calculated the
distance  of  V4444~Sgr  as $d_\text{c}>5$  kpc.   \citet{Venturinie02}  also
estimated the reddening  as $E(B-V) = 0.48 \pm0.15$  mag comparing the
observed ratios of the hydrogen lines to Pa$\beta$, corresponding to a
distance of $d_\text{c}\sim2.3$  kpc.  Using the field stars  near V4444 Sgr,
\citet{Kawabatae00}  estimated the  colour excess  as $E(B-V)\sim0.75$
mag.  For this estimate, our reddening  curve implies a lower limit of
the distance as  $d_\text{c}\ge3.7$ kpc.  We also calculated  the distance to
be $d_\text{c}=2.2$  kpc using a reddening  estimate of $E(B-V) =  0.431$ mag
\citep{Kamathe08}  from   the  open  cluster  NGC~6520   located  near
V4444~Sgr.  We can  analyse the reddening evolution  through this nova
only  for  reddening  values   $E(B-V)\leq0.7$  mag  corresponding  to
distances smaller  than 5  kpc, due  to  the magnitude  depth of
2MASS and the dwarf/giant  contamination.  Since the reddening becomes
constant  for  distances larger  than  5  kpc,  we conclude  that  the
reddening estimate  given by \citet{Venturinie02}  based on O ~\textsc{i} lines
and \citet{Kawabatae00} are possibly overestimated, or the distance is
larger  than  5 kpc.   Thus,  using  an  average  value of  $E(B-V)  =
0.45\pm0.1$  mag,   we  obtained   the  distance   of  this   nova  as
$d_\text{c}=2.27\pm0.58$ kpc.
\subsubsection{V4633 Sgr}
To   estimate   the   interstellar  reddening   towards   V4633   Sgr,
\citet{Lynche01}   used   hydrogen   lines  and   determined   it   as
$E(B-V)=0.28\pm0.13$ mag, while the He \textsc{ii} lines yielded values between
0.13-0.49 mag.  Combining these measurements, \citet{Lynche01} adopted
$E(B-V)=0.3\pm0.2$ mag,  but also mentioned that  the uncertainties in
the reddening were dominated by  errors in the measured line strengths
because of blending, complex  line profiles, and the signal--to--noise
ratio  of   observations.  Additionally,   \citet{Lipkine01}
investigated the  reddening towards  V4633 Sgr using  several methods:
$E(B-V)=0.21\pm0.03$   mag   from   analyses  of   Balmer   decrement,
$E(B-V)\sim0.23$ mag  from the  He triplet  ratio, $\lambda$5876/4471,
$E(B-V)\leq0.25$ mag from comparison  between intrinsic colours at max
and $t_2$.  We used the simple arithmetic mean of these results,
$E(B-V)=0.26\pm0.05$    mag,   and    estimated   the    distance   as
$d_\text{c}=1.36\pm0.46$ kpc.
\subsubsection{V4643 Sgr}
In order to calculate the distance of V4663 Sgr, we used the reddening
$E(B-V)=1.47\pm0.2$  mag  \citep{Ashoke06}  estimated  from  intrinsic
colour, and determined the distance as $d_\text{c}=3.10\pm0.20$ kpc.
\subsubsection{V5113 Sgr}
In  this  paper, the  distance  of  V5113  Sgr  was calculated  to  be
$d_\text{c}=0.95\pm0.21$  kpc  using   $E(B-V)=0.1\pm0.02$  mag  reported  by
\citet{Ruch04},   who  derived   the   reddening   from  the   diffuse
interstellar line at 5780 {\AA}.
\subsubsection{V5115 Sgr}
Using   the  reddening   estimate  of   $E(B-V)=0.53\pm0.05$  mag   by
\citet{Rudye05}, whose  determination is  based on the  O~\textsc{i}  lines, we
calculated the distance to be $d_\text{c}\sim3.0\pm1.0$ kpc.
\subsubsection{V5116 Sgr}
V5116 Sgr (Nova Sagittarii 2005 No. 2) was discovered on July 4.049 UT
(JD 2453555.549)  at about  $V=8.0$ mag \citep{Liller05}  and observed
also as  a supersoft  X-ray source \citep{Salae08}.   Fortunately, the
time  and  magnitude  of  the outburst  maximum  are  reasonably  well
determined for V5116 Sgr, with  a pre-maximum observation 24 hr before
the maximum.  The colour at  the maximum as $B-V=+0.48$ given by
\citet{GilmoreK05}  indicate  that the  interstellar  reddening
towards 5116 Sgr as $E(B-V)=0.25\pm0.06$  mag using intrinsic colour of
$(B-V)_0=0.23\pm0.06$    \citep{vandenbergY87}.    This    result   is
consistent with $E(B-V)=0.22\pm0.02$ mag \citep{Salae10} obtained from
the column  density of $N_H=1.3\pm0.1 \times  10^{21}$ cm$^{-2}$ using
$N_H=5.9\times10^{21} \ E(B-V)$  cm$^{-2}$ \citep{Zombeck07}, and also
in an agreement with $E(B-V)=0.24\pm0.08$ mag given by \citet{Nesse07}
from SWIFT  X-ray observations.  In  this paper, we  therefore adopted
$E(B-V)=0.23\pm0.06$    mag   and    calculated   the    distance   as
$d_\text{c}=1.55\pm0.70$ kpc.
\subsubsection{V5117 Sgr}
V5117  Sgr   is  a  standard  Fe   \textsc{ii}  nova.  The  colour   excess  of
$E(B-V)=0.5\pm0.1$   mag    was   obtained    from   the    OI   lines
\citep{Lynche06}.   Using this  value, we  calculated its  distance as
$d_\text{c}=1.45\pm0.33$ kpc.
\subsubsection{V5558 Sgr}
\citet{Munarie07a} obtained $E(B-V) =  0.36$ mag from the interstellar
Na~\textsc{i}~D1 and  D2 lines, while \citet{Rudye07}  determined the reddening
of $E(B-V )\sim  0.8$ mag based on  the relative strengths of  the O \textsc{i}
$\lambda$8446  and $\lambda$12187  lines.  Although  both methods  are
known to  yield reliable  estimates of the  reddening, they  give very
different results.   Using these reddening estimates,  the distance of
the nova were determined between $1.1\pm0.2$ and $2.1\pm0.4$ kpc, with
a  mean of  $d_\text{c}=1.6$ kpc.   In addition,  \citet{Hachisu14} concluded
that  the  reddening  should  be  $E(B-V)=0.7$  mag,  consistent  with
\citet{Rudye07}, assuming that the three novae, V5558 Sgr, HR Del, and
V723 Cas, have the same  intrinsic $(B-V)_0$ colour in the pre-maximum
phase.  This  approach implies  that the  best reddening  estimate for
V5558 Sgr is the one given by \citep{Rudye07}.
\subsubsection{V732 Sgr}
\citet{Weight93} used  the relation between  column density of  CO and
extinction, and  estimated the reddening as  $E(B-V)=0.81\pm0.16$ mag,
indicating a distance of $d_\text{c}=3.0\pm0.2$ kpc in our study.
\subsubsection{CK Vul}
\citet{Hajduketal13}   determined   the  interstellar   reddening   as
$E(B-V)=0.7$  mag for  this system,  which is  close to  the reddening
$E(B-V)=0.82\pm0.23$ mag  given by \citet{Sharaetal85}.   These values
were inferred  from the  Balmer decrement.  \citet{Orio01}  used ROSAT
observation  and   calculated  $E(B-V)=0.73$   mag.   \citet{Weight93}
analysed  emission  lines  associated  with  CO,  and  determined  the
reddening as $E(B-V)=0.8$ mag which  is consistent with that estimated
by  \citet{Sharaetal85}   within  errors.   Since  all   the  reliable
spectroscopic measurements are in  agreement, we adopted the reddening
of  CK~Vul  as  the  arithmetic  mean  $E(B-V)=0.75\pm0.05$  mag,  and
calculated the distance of this nova as $d_\text{c}=4.48\pm0.24$ kpc.
%
\section{The novae which their distance could not be calculated.}
This  section  includes  a  table of  novae  and  their  adopted
  reddening  estimates  in  the   literature  together  with  a  short
  explanation of why  their distances could not be  calculated in this
  study.
\begin{table*}
\label{notcalc}
\begin{center}
\caption{Novae for which the distance could not be calculated}
	\begin{tabular}{lccccclccccc}
\hline\hline
Name  & $l\degrees$ & $b\degrees$ & $E(B-V)$ & Ref$^a$ & Notes$^b$ & Name  & $l\degrees$ & $b\degrees$ & $E(B-V)$ & Ref$^a$ & Notes$^b$  \\
\hline
OS And   & 106.052 & -12.118 & 0.15 & 1 &   l   & GI Mon   & 222.93 & 4.749 &  $0.10\pm0.04$  & 8 &   l, c:\\
V723 Cas   & 124.961 & -8.807 & 0.45 & 2 &   c, e:   & RS Oph   & 19.799 & 10.372 &  $0.73\pm0.1$  & 12 &   c \\
V693 CrA   & 357.83 & -14.391 &  $0.05\pm0.05$  & 3 &   l  & V841 Oph   & 7.621 & 17.778 &  $0.44\pm0.06$  & 8 &   s, c:\\
T CrB   & 42.374 & 48.165 &  $0.1\pm0.1$  & 4 &   l  & V849 Oph   & 39.233 & 13.48 & 0.27 & 13 &   l \\
V1016 Cyg   & 75.173 & 5.678 & 0.28 & 5 &   s, c: & AG Peg   & 69.278 & -30.887 &  $0.12\pm0.04$  & 14 &   l\\
V1668 Cyg   & 90.837 & -6.76 &  $0.38\pm0.08$  & 6 &   c & GK Per   & 150.956 & -10.104 &  $0.34\pm0.04$  & 8 &   s \\
V1819 Cyg   & 71.372 & 3.978 &  $0.35\pm0.15$  & 7 &   e  & RR Pic   & 272.355 & -25.672 &  $0.00\pm0.02$  & 8 &   l \\
HR Del   & 63.43 & -13.972 &  $0.17\pm0.02$  & 8 &   l  & T Pyx   & 257.207 & 9.707 &  $0.25\pm0.2$  & 15 &   c \\
V339 Del   & 62.199 & -9.423 &  $0.18\pm0.035$  & 9 &   l  & T Sco   & 352.675 & 19.462 & 0.19 & 16 &   c\\
KT Eri   & 207.986 & -32.02 & 0.08 & 10 &   l  & U Sco   & 357.669 & 21.869 & 0.15 & 17 &   c\\
DN Gem   & 184.018 & 14.714 &  $0.17\pm0.04$  & 8 &   l, s:  & X Ser   & 10.841 & 31.872 & 0.25 & 18 &   l\\
DQ Her   & 73.153 & 26.444 &  $0.05\pm0.02$  & 8 &   l  & V4743 Sgr   & 14.124 & -11.877 & 0.25 & 19 &   s\\
V533 Her   & 69.189 & 24.273 &  $0.03\pm0.02$  & 8 &   l & V4745 Sgr   & 1.493 & -12.382 & 0.1 & 20 &   l, c:\\
V827 Her   & 45.808 & 8.594 &  $0.44\pm0.06$  & 8 &   c & V5114 Sgr   & 3.943 & -6.312 & 0.45 & 21 &   c, e: \\
DI Lac   & 103.108 & -4.855 &  $0.26\pm0.04$  & 8 &   e  & RR Tel   & 342.163 & -32.242 & 0.0-0.1 & 22 &   l\\
GW Lib   & 340.707 & 26.767 & 0.03 & 11 &   l & PU Vul   & 62.575 & -8.532 &  $0.43\pm0.05$  & 23 &   e\\
HR Lyr   & 59.584 & 12.47 &  $0.18\pm0.06$  & 8 &   l, c:  &  &  &  &  &  & \\
\hline\hline
\end{tabular}
\end{center}
\begin{flushleft} \footnotesize{
$^a$ References for reddening estimates: 
1 -- \citet{Hachisu14};
2 -- \citet{Munarie96};
3 -- \citet{HachisuK16b};
4 -- \citet{Schaefer10};
5 -- \citet{Arkhipovae08};
6 -- \citet{SlovakV79};
7 -- \citet{WhitneyC89};
8 -- \citet{SelvelliG13};
9 -- \citet{Chochole14};
10 -- \citet{Ragane09};
11 -- \citet{Bullocke11};
12 -- \citet{Snijders87};
13 -- \citet{Szkody};
14 -- \citet{PenstonA85};
15 -- \citet{GilmozziS07};
16 -- \citet{Cohen85};
17 -- \citet{Mason11};
18 -- \citet{Selvelli04};
19 -- \citet{Vanlandinghame07};
20 -- \citet{Csake05};
21 -- \citet{Ederoclitee06};
22 -- \citet{Younge05,Selvellie07};
23 -- \citet{LunaC05} \\
$^b$ 
c -- obtained reddening curve is nearly constant within the magnitude limits.;
e -- although the reddening slightly increases with distance, the reddening estimate remains within the error limits of each reddening point in relation.;
l -- the interstellar reddening through line of sight is too low to obtain a smooth reddening curve. ;
s -- position of the RC stars on CMD are unclear, since there are only a few of them. ;
: --  secondary reason, but its efficacy is lesser. 
}
\end{flushleft} 
\end{table*}	
\section{The Reddening-Distance Relations}
\label{appBsec}
In this section, the reddening-distance relations, from which the distances of novae are estimated
in this study, are given. 
 	\begin{figure*}
	\centering
	\includegraphics[width=0.8\textwidth]{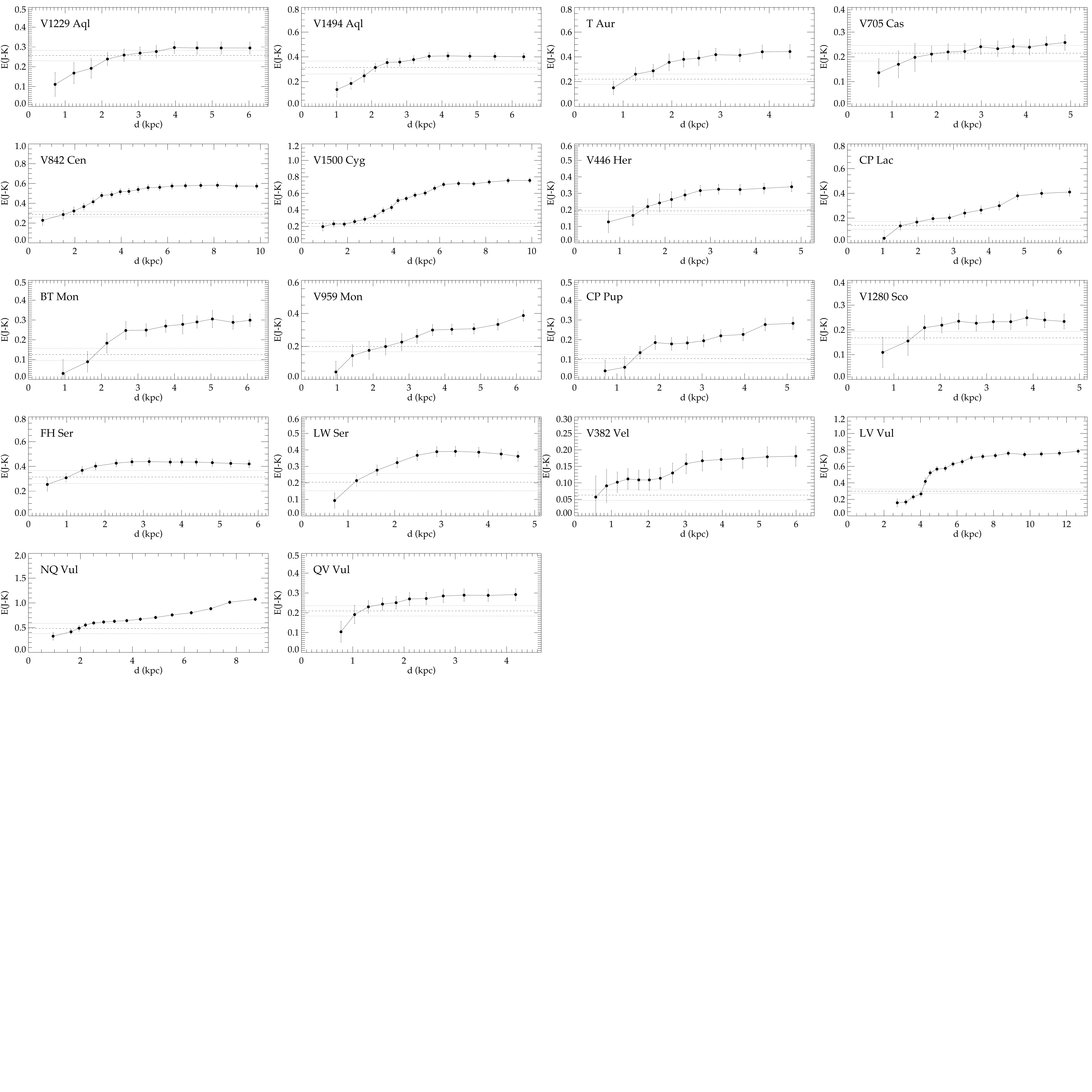}
	\caption{The reddening-distance  relations for  the directions
          of  the  novae,  for  which expansion  parallaxes  could  be
          measured. These novae are listed in Table~\ref{table:exp}.}
	\label{Fig:A1}
	\end {figure*}

 	\begin{figure*}
	\centering
	\includegraphics[width=0.8\textwidth]{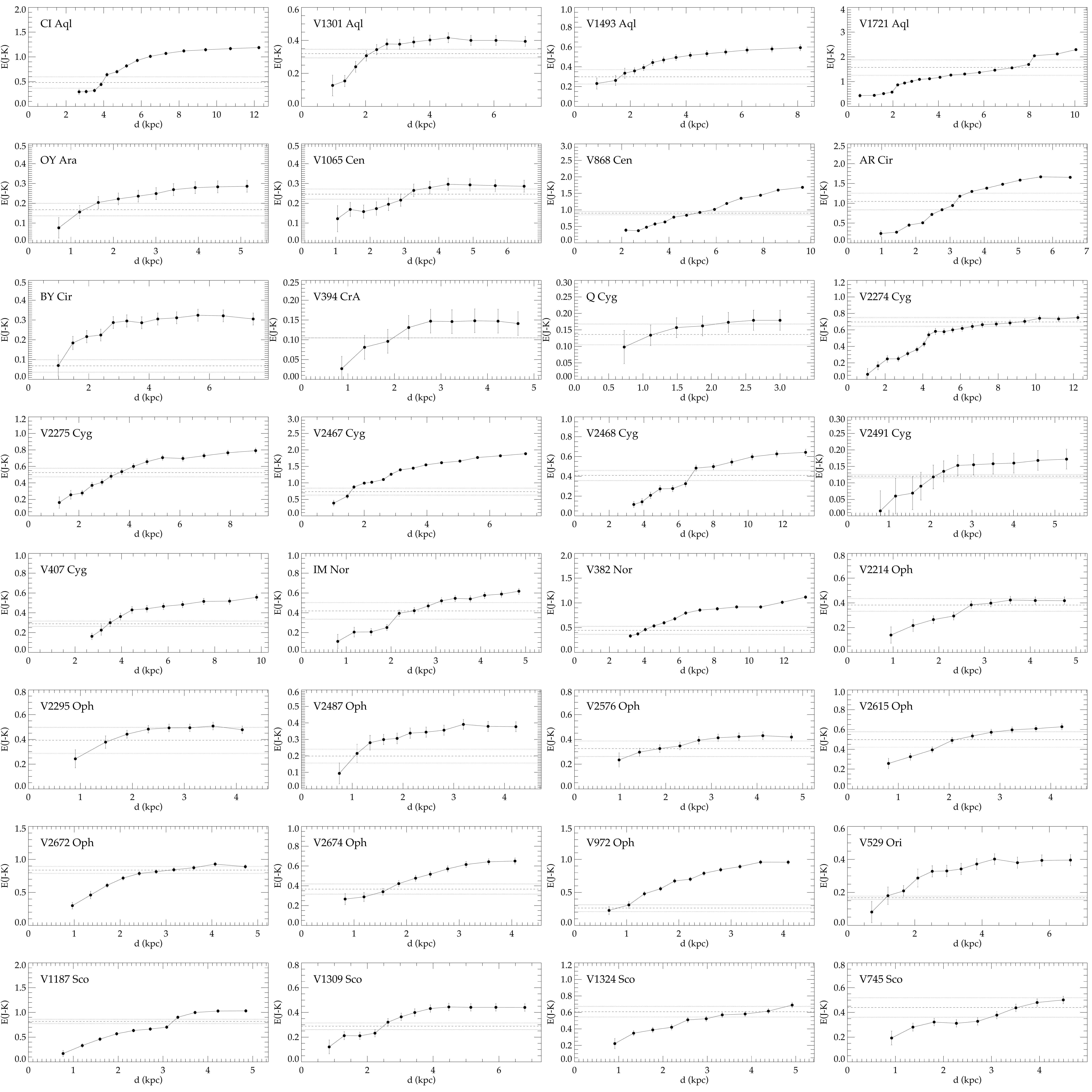}
	\caption{cont.}
	\label{Fig:A2}
	\end {figure*}
	
\addtocounter{figure}{-1}
	 \begin{figure*}
	\centering
	\includegraphics[width=0.8\textwidth]{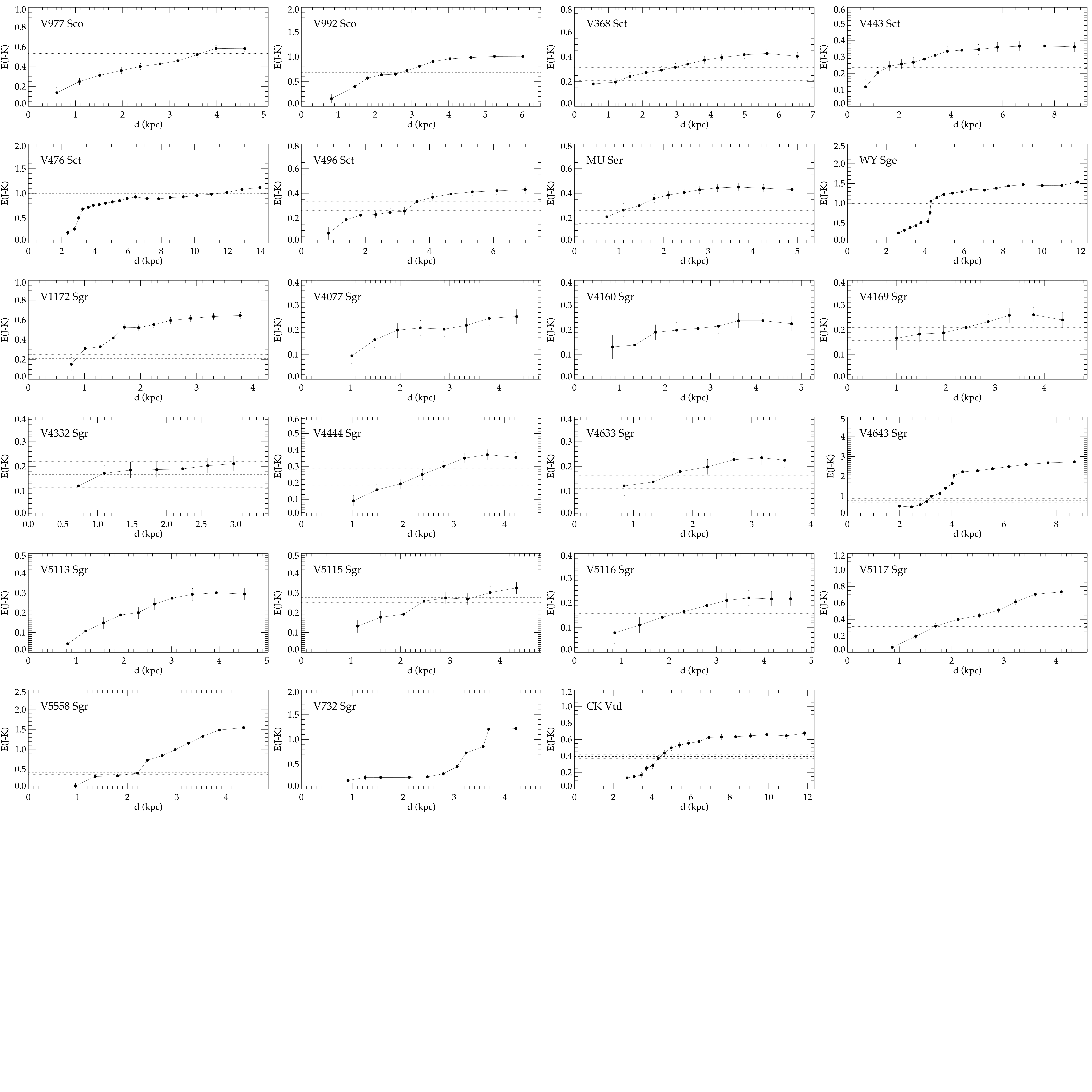}
	\caption{The reddening-distance relations for novae listed in Table~\ref{table:wexp}}
	\label{Fig:A3}
	\end {figure*}
	
	 \begin{figure*}
	\centering
	\includegraphics[width=0.8\textwidth]{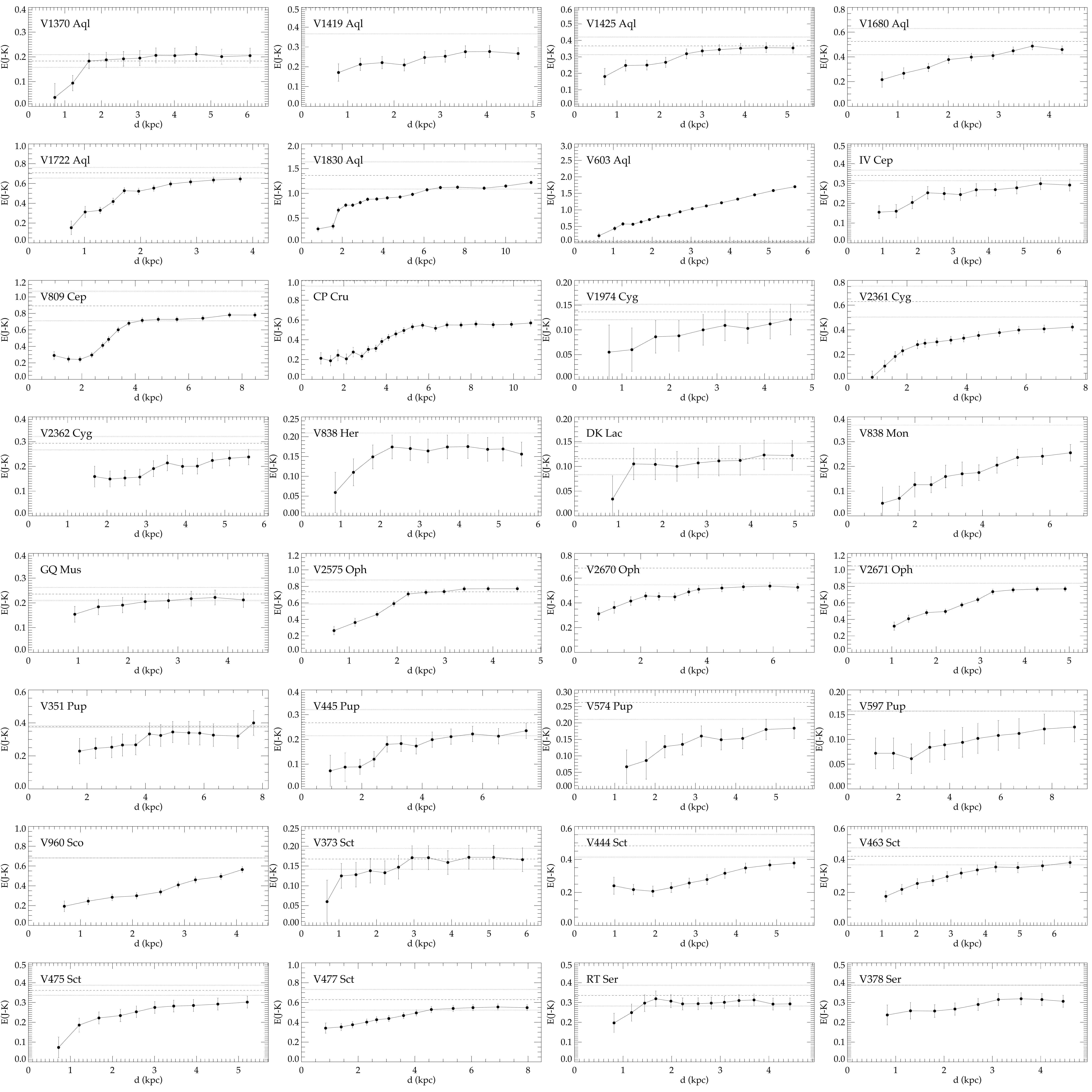}
	\includegraphics[width=0.8\textwidth]{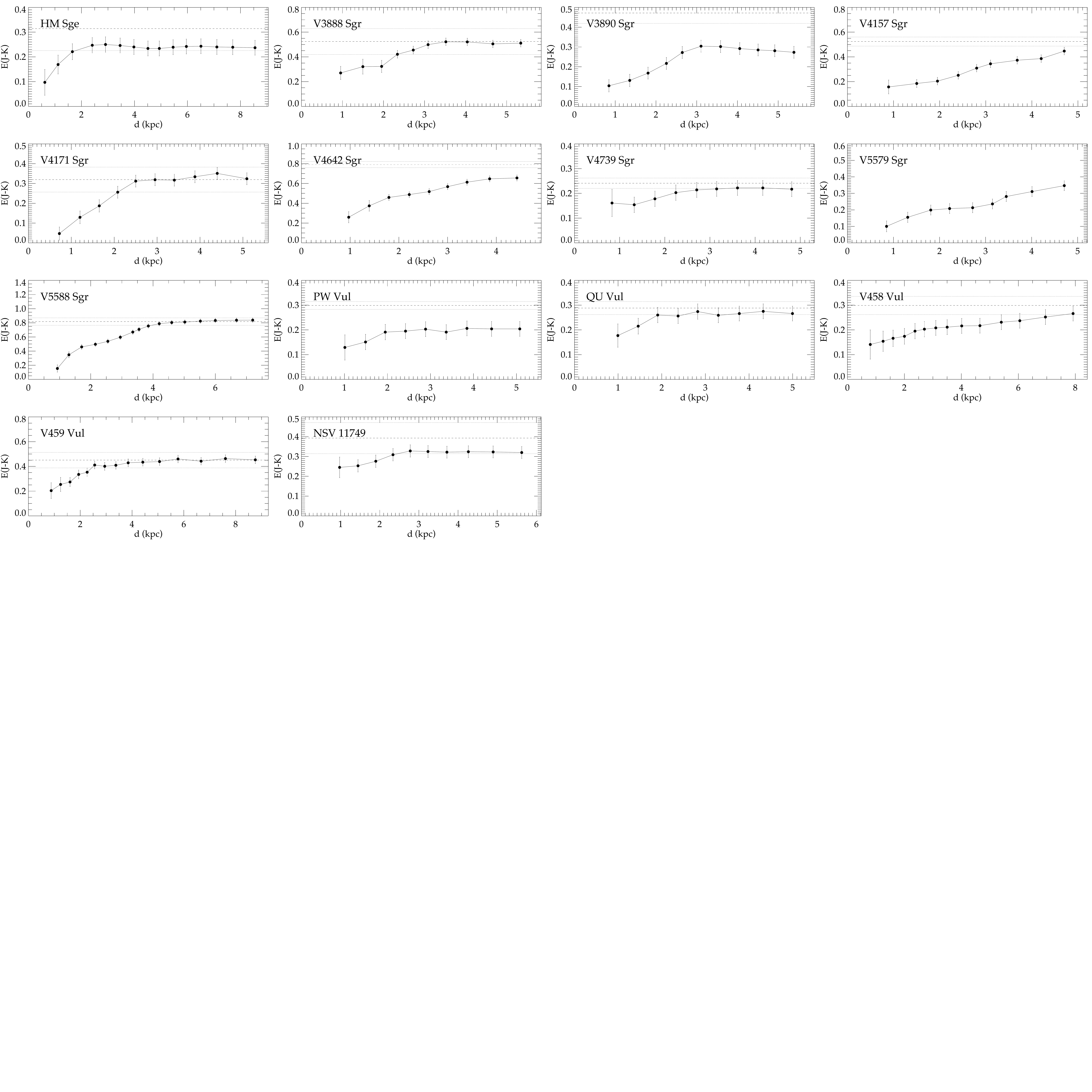}
	\caption{The reddening-distance relations for novae listed in Table~\ref{table:lowliml}}
	\label{Fig:A5}
	\end {figure*}
\bsp
\label{lastpage}
\end{document}